\documentclass{article}
\usepackage{PRIMEarxiv}
\usepackage[utf8]{inputenc} 
\usepackage[T1]{fontenc}    
\usepackage{hyperref}       
\usepackage{url}            
\usepackage{booktabs}
\usepackage{amssymb}       
\usepackage{amsfonts}       
\usepackage{nicefrac}       
\usepackage{microtype}      
\usepackage{lipsum}
\usepackage{fancyhdr}       
\usepackage{graphicx}       
\graphicspath{{media/}}     
\usepackage{natbib}
\usepackage{algorithm}
\usepackage{algpseudocode}
\usepackage{amsfonts,amsmath,amssymb}
\usepackage{cancel}
\usepackage{float}
\usepackage{mdframed}
\usepackage{subfig}

\newcommand{\dif}{\ensuremath{\mathrm{d}}}

\newcommand{\T}{\ensuremath{\mathrm{\scriptscriptstyle T}}}
\usepackage{bm}

\newcommand{\betab}{\ensuremath{\bm\beta}}

\usepackage{amsmath}

\DeclareMathOperator{\E}{E}
\DeclareMathOperator{\var}{var}

\DeclareMathOperator{\iid}{\overset{iid}{\sim}}

\usepackage{mathrsfs}

\newlength\aftertitskip     \newlength\beforetitskip
\newlength\interauthorskip  \newlength\aftermaketitskip

\if@abbrvbib
\else
\fi

\bibpunct{(}{)}{;}{a}{,}{,}

\newtheorem{theorem}{Theorem}
\newtheorem{definition}{Definition}
\newtheorem{example}{Example}
\newtheorem{proposition}{Proposition}

\newtheorem{lemma}{Lemma}

\usepackage{xcolor}

\def\gray#1{{\color{gray} #1}}

\newcommand{\restr}[1]{|_{#1}}
\def\Xint#1{\mathchoice
{\XXint\displaystyle\textstyle{#1}}%
{\XXint\textstyle\scriptstyle{#1}}%
{\XXint\scriptstyle\scriptscriptstyle{#1}}%
{\XXint\scriptscriptstyle\scriptscriptstyle{#1}}%
\!\int}
\def\XXint#1#2#3{{\setbox0=\hbox{$#1{#2#3}{\int}$ }
\vcenter{\hbox{$#2#3$ }}\kern-.6\wd0}}

\def\dashint{\Xint-}

\title{Uncovering Regions of Maximum Dissimilarity \\ on Random Process Data}

\author{
  Miguel de Carvalho\\
  School of Mathematics\\
  University of Edinburgh\\
  EH93FD, Edinburgh, UK\\
  \texttt{miguel.decarvalho@ed.ac.uk} \\
   \And
  Gabriel Martos\\
  Departamento de Matemática y Estadística\\
  Universidad Torcuato Di Tella\\
  Buenos Aires, Argentina\\
  \texttt{gmartos@utdt.edu} \\
}

\begin{document}
\maketitle

\begin{abstract}
    The comparison of local characteristics of two random processes can
  shed light on periods of time or space at which the processes
  differ the most.  This paper proposes a method that learns about
  regions with a certain volume, where the marginal attributes of two
  processes are less similar. The proposed methods are devised in full
  generality for the setting where the data of interest are
  themselves stochastic processes, and thus the proposed method can
  be used for pointing out the regions of maximum dissimilarity with a
  certain volume, in the contexts of functional data, time series, and
  point processes.  The parameter functions underlying both stochastic
  processes of interest are modeled via a basis representation, and
  Bayesian inference is conducted via an integrated nested Laplace
  approximation. The numerical studies validate the proposed
  methods, and we showcase their application with case studies on
  criminology, finance, and medicine.
\end{abstract}


\section{Introduction}\label{intro}
Everyday millions of data patterns flow around the world at
unprecedented speed, thus leading to an explosion on the demand for
modeling stochastic process data---such as time series, point
processes, and functional data; each of these types of data plays a key
role in machine learning, as can be seen, for
instance, from the recent papers of \cite{berrendero2020},
\cite{faouzi2020}, and \cite{xu2020}.  Hand in hand with this shock on
demand arrived a pressing need for the development of data-intensive
methods, techniques, and algorithms for \textit{learning and comparing
random processes}. 



\subsection{The Learning Problem of Interest}\label{problem}
In this paper we deal with the following problem on the comparison of stochastic processes: 
\begin{quote}
\textsc{learning problem}.~\textit{For a pair of random processes, what is 
the region---with a given volume---where they statistically differ the most?}
\end{quote}

\noindent Specifically, the task of interest will entail tracking down
regions with a certain volume, where the marginal attributes of two
stochastic processes differ the most. Throughout, we will refer to
this problem as that of uncovering \textit{regions of maximum
  discrimination} (RMD) on random process data.

Let's translate the learning problem into mathematical parlance. 
Since the target of interest consists of a set that fulfills an
optimization criterion---i.e.~a period of time or region over which a
marginal feature of two stochastic processes differ the most---some of the
main concepts in this paper can be framed as an optimization problem over
a set function. The canonical problem in set function optimization 
is of the type, 
\begin{equation}\label{setopt}
\begin{array}{rl}
  \underset{A \subseteq \mathscr{A}}{\max} &F(A) \\
  \text{s.t.} & A \in \mathscr{F},
\end{array}
\end{equation}
where $F: 2^{\mathscr{A}} \rightarrow \mathbf{R}$ is a set function,
$\mathscr{A}$ is a set, and $\mathscr{F} \subseteq 2^{\mathscr{A}}$ is
the collection of feasible sets defining the constraint.  Optimization
problems over set functions---such as \eqref{setopt}---are commonplace
in machine learning \citep[e.g.][]{krause2010}.  For
a recent review on the theory of discrete set function optimization see \cite{wu2019}.
Most state of the art developments have been made on \textit{discrete}
or \textit{combinatorial set function optimization}, especially on the
class of submodular functions \citep[e.g.][]{goldengorin2009}.  Our
paper provides one of the first steps towards \textit{continuous set
  function optimization} as here the aim will be to solve \eqref{setopt}
when $\mathscr{F}$ is a family of Borel subsets of a compact set $T \subset \mathbf{R}^d$.

As it will be seen in Section~\ref{model}, the objective set function
of interest in our case will be a measure of proximity between
marginal features of the pair of stochastic processes of interest,
whereas the collection of feasible sets introduces the constraint on
the `size' of the feasible regions---i.e. periods of time or space---over which the comparison is made.

\subsection{Our Contributions}\label{contributions}
Our main contributions are:
\begin{enumerate}
\item We pioneer the study, formulation, and analysis of the learning
  problem of tracking down regions of maximum discrimination as described
  in Section~\ref{problem}---and formally defined in
  Sections~\ref{model}--\ref{extensions}.
\item In its most standard version, the proposed learning problem is shown to be equivalent to a continuous set
function optimization on a monotone modular function, under a Lebesgue
measure constraint. Hence, as a byproduct, the paper contributes to the
literature on set function optimization which is mostly focused on a
discrete and combinatorial framework, with a particular emphasis on 
monotone submodular functions \cite[e.g.][]{nemhauser1978,
  calinescu2011, goldengorin2009, buchbinder2017, buchbinder2018},
under cardinality or matroid constraints. Little is known
  on continuous set function optimization, and thus the tools, concepts, and strategies 
  devised herein can be of further interest elsewhere.
\item Our approach is fully general in the sense that it applies to
  most random process data (including functional data, time series,
  and point processes). The functional
  parameters of the processes of interest (say mean functions,
  volatility functions, or intensity functions) are modelled by
  composing an inverse link function with a basis function
  representation, and Bayesian modeling is conducted via
  latent Gaussian models and inference
  is conducted via INLA (Integrated Nested Laplace Approximations)
  \citep{rue2009, rue2017}.
\item An extension of the proposed method applies also to the context where the interest is on
  comparing more than one marginal feature via a multi-objective
  version of the set function optimization problem of interest.
\item A variant of the proposed approach shows that 
  a well-known performance measure known as Youden index and the Youden's $\text{J}$ statistic
  \citep[e.g.][]{sokolova2006, inaciodecarvalho2017} as well as the Kolmogorov metric \citep[e.g.,][]{gretton2012kernel} can be regarded as particular cases of the framework developed herein.  
\end{enumerate}

\subsection{Some Related Prior Work}\label{previous}
While the learning problem introduced in Section~\ref{problem} is
new---and while there are novel contributions that will arise from our
solution to it---there are some recent approaches in the context of
functional data analysis \citep{ramsay2002, ramsay2005, ferraty2006,
  horvath2012} that are tangentially related to
it, and that are briefly reviewed below. 

\cite{pini2016, pini2017} propose an interval
testing procedure for functional data that points out specific
differences between functional populations. Also in the context of
functional data, \cite{berrendero2016} propose a discretization
method consisting on learning to choose a finite collection of points
in the domain of a set of functions in order to improve the
performance of a functional data classifier. \cite{martos2018} propose
a Mann--Whitney type of statistic for functional data so to learn
about the regions at which two processes differ the most on
aspects related with symmetry. Finally, \cite{dette2020} develop 
hypothesis tests for the equivalence of functional parameters in a two
sample functional data setup.

Our approach differs from the ones mentioned above in a number of
important ways. Perhaps the most important ones are that: \textit{i}) here the
goal is not to test hypothesis but rather to learn about regions with
a given volume where two processes differ the most; \textit{ii}) our approach
applies to random processes in general whereas the methodologies
reviewed above have mainly been designed with the context of functional data analysis in mind.

\subsection{Outline of the Paper}
The rest of the paper unfolds as follows. 
In Section~\ref{model} we introduce
sets of maximum dissimilarity, present examples, and introduce the
inference methods. In Section~\ref{extensions} we comment on extensions of
the main concepts, methods, and ideas covered in Section~\ref{model}. 
In Section~\ref{simulation} we report the main findings of
a Monte Carlo numerical study on artificial data. In Section~\ref{data} we showcase the
proposed methods on real data applications. Closing remarks are given in
Section~\ref{discussion}. Appendix~\ref{lemmata} includes a selection of
auxiliary facts, and the proofs of main results are included in
Appendix~\ref{proofs}. Table~\ref{symbols} lists symbols and notation used throughout the article.
We write $X(t)$ instead of $X_t$ when typographically convenient.

\begin{table}[H] 
  \centering
  \begin{tabular}{ll}
    \hline    
    \textbf{Symbol} & \textbf{Description} \\
    \hline 
    $X$, $Y$ & stochastic processes under comparison, that is, $X = \{X_t\}$ and $Y = \{Y_t\}$ \\
    $T$ & ground, or index, set over which processes $X$ and $Y$ are defined \\
    $\theta_{X}$, $\theta_{Y}$ & functional parameters for $X$ and $Y$ \\
    $\mathcal{D}_X, \mathcal{D}_Y$ & data on proceses $X$ and $Y$ \\
    $\|\cdot\|^{(A)}_p$ & $L^p$ sub-norm over $A$\\
    $F(\cdot)$ & set objective function $\|\theta_X - \theta_Y\|^{(\cdot)}_p$ \\
    $f(t, r)$ & set objective function evaluated at closed ball $B(t, r)$, that is, $F\{B(t, r)\}$ \\
    $F_i$ & set objective functions in multi-objective context \\
    $\mathcal{A}$ & set of compact and convex subsets of the ground set $T$ \\
    $|A|$ & volume functional, i.e., Lebesgue measure of $A$ in $\mathbf{R}^d$ \\
    $\mathscr{F}_c$ & collection of feasible sets \\
    $A^*_c$ & region of maximum or multi-maximum dissimilarity \\
    $D^*_c$ & dissimilarity index for region of maximum dissimilarity \\
    $B^*_c$ & ball of maximum or multi-maximum dissimilarity \\
    $\mathscr{D}^*_c$ & dissimilarity index for ball of maximum dissimilarity \\
    $t_{c}^*, r^*_c$ & center and radius of ball of maximum dissimilarity \\
    $\mathscr{B}_p$ & set of all closed balls in $L^p$ \\
    $d_{\mathrm {H} }(A, B)$ & Hausdorf distance between sets $A$ and $B$ \\
    $d(x, B)$ & minimum Euclidean distance between a point $x \in A$ and set $B$ \\
    $g\restr{A}$ & restriction of function $g$ to set $A$ \\
    $\partial A$ & boundary of set $A$ \\
    $\dashint$ & average integral symbol \\
    $\twoheadrightarrow$ & notation for defining correspondences (i.e.~set-valued functions) \\
    \hline    
  \end{tabular}  
  \caption{\label{symbols} Symbols and notation used throughout the article}
\end{table}
\section{Learning about Sets of Maximum Dissimilarity}\label{model}
\subsection{Sets of Maximum Dissimilarity}\label{sec:dis}
\subsubsection*{A taster---some simple instances}
\begin{figure}
  \centering
  \begin{tabular}{ccc}
\subfloat[]{\includegraphics[scale=0.18]{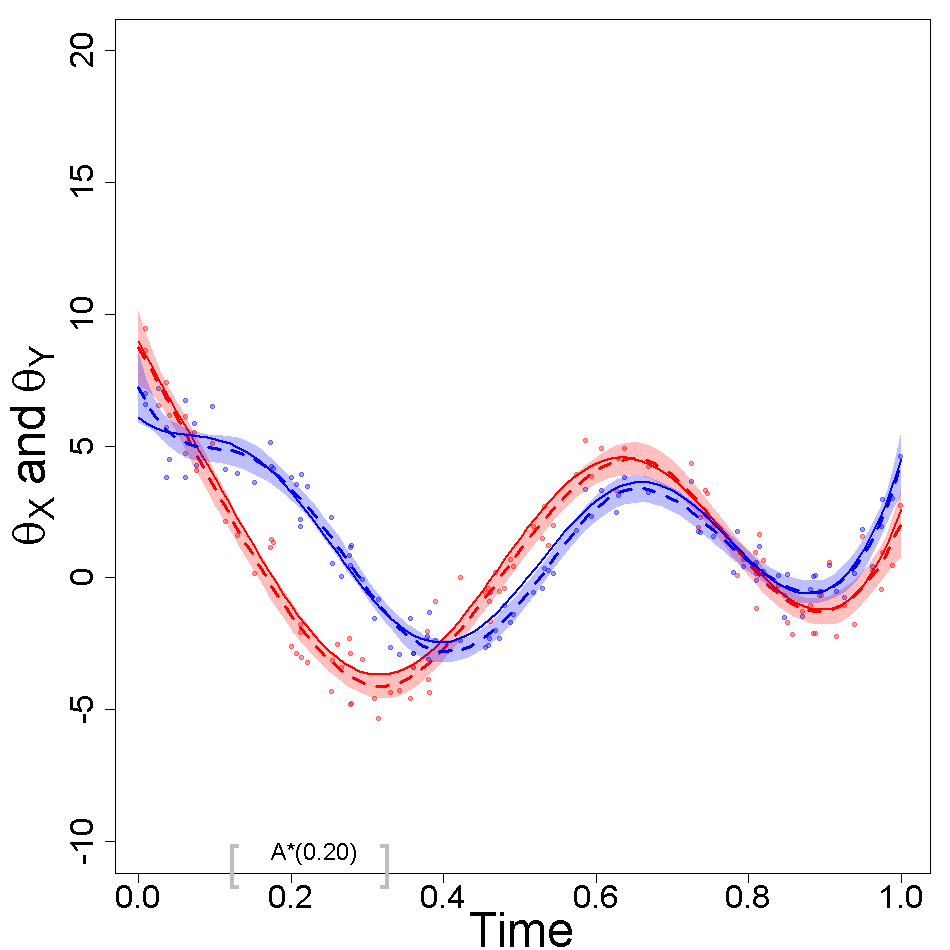}
} &
\subfloat[]{\includegraphics[scale=0.18]{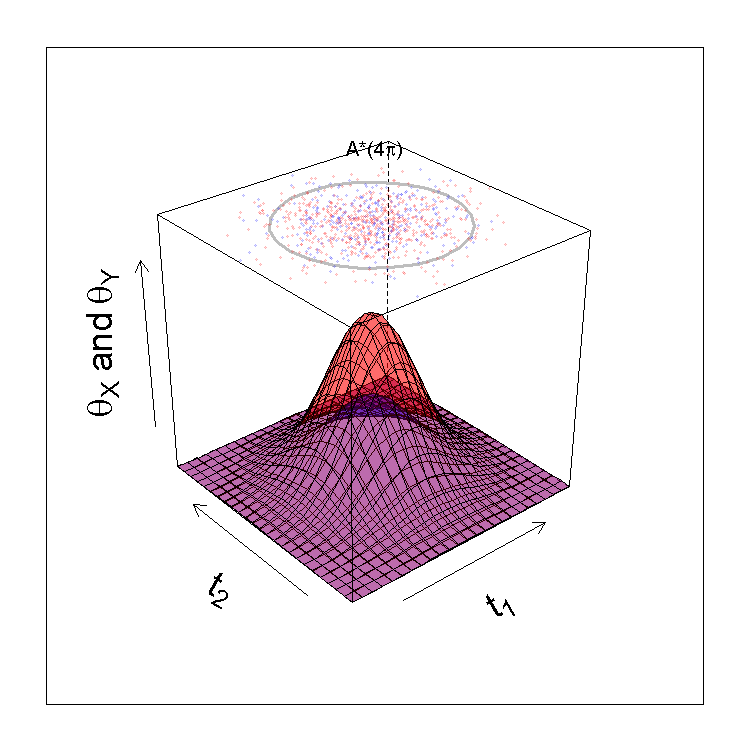}
}&
\subfloat[]{\includegraphics[scale=0.18]{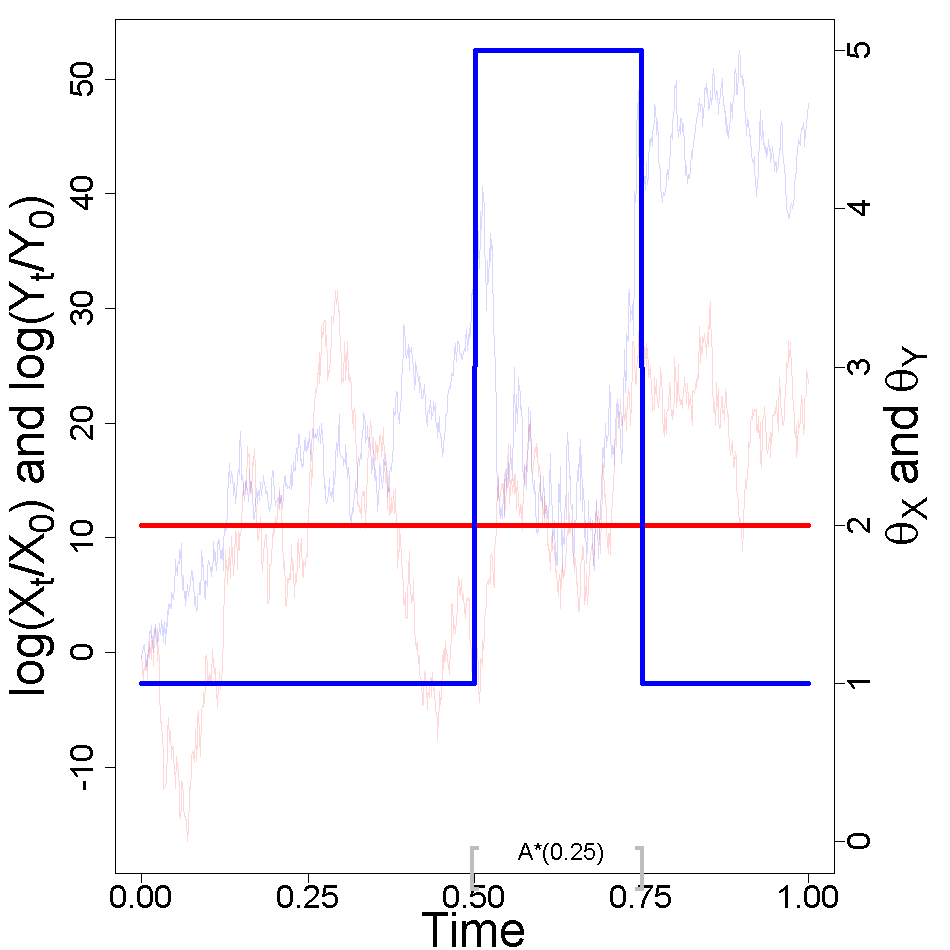}
}
\end{tabular}
\caption{Sets of maximum dissimilarity for Examples~\ref{ex:1}--\ref{ex:3}. The sets are depicted using brackets in the time axis  in the case of mean functions and volatility functions (a, c) and on the top of the box with a line in the case of intensity functions (b). Raw data alongside the true and estimated functional parameters $\theta_{X}$ and $\theta_{Y}$ are depicted in red  and blue respectively; the proposed inferences for the sets will be discussed in Section~\ref{learningfromdata}.
  \label{fig:1}}
\end{figure}

\noindent Prior to introducing sets of maximum dissimilarity in a formal fashion, we introduce some simple instances of the concept based on different parameter functions $\theta_{X}$ and $\theta_{Y}$. 
The respective sets of maximum dissimilarity and parameter functions for Examples~\ref{ex:1}--\ref{ex:3} to be presented below are depicted in Fig.~\ref{fig:1}. 

\begin{example}[Mean Functions]\label{ex:1}   \normalfont 
  Let $X_t = \theta_{X}(t) + \varepsilon_X(t)$ and  $Y_t = \theta_{Y}(t) + \varepsilon_Y(t)$ be two stochastic processes defined on the unit interval, $T=[0,1]$, with 
  \begin{equation}\label{funex1}
    \begin{cases}
      \theta_{X}(t) = E(X_t) = b(t) + 4\cos(10t) - 2(t-0.75)^2, \\
      \theta_{Y}(t) = E(Y_t) = b(t)  + 3\sin(12t), 
    \end{cases}    
  \end{equation}
   where $b(t) = 1 / 2 \exp\{10(t-0.5)^2\}$ is a baseline curve, and $\varepsilon_X(t)$ and $\varepsilon_Y(t)$ are zero mean Gaussian error functions. Fig.~\ref{fig:1}(a) shows the connected set of size $c=0.2$ where both mean functions differ the most in the $L^1$ sense.
\end{example}

\begin{example}[Intensity Functions\label{ex:2}] \normalfont 
  Consider the intensity functions associated to two point processes 
  \begin{equation}
    \begin{cases}\label{funex2}
    \theta_{X}(t) = \lambda_X(t) = \gamma\exp\{ -(t_1^2 +t_2^2)/2  \}, \\
    \theta_{Y}(t) = \lambda_Y(t) = \delta \theta_{X}(t),      
    \end{cases}
  \end{equation}
  defined over the region $T = [-3, 3]^{2}$ with $\gamma, \delta > 0$. 
Fig.~\ref{fig:1}(b) shows the true ball of maximum dissimilarity with area $4\pi$ at which the two intensity functions differ the most in the $L^2$ sense in the case where $\gamma =100$ and $\delta = 1 / 2$.
\end{example}

\begin{example}[Volatility Functions]\label{ex:3} 
  \normalfont Suppose
  $X_t$ and $Y_t$ are the log returns of two stock markets, with $E(X_t) = E(Y_t) = 0$, and that the
  goal is to search for the period of about a quarter (think of
  $T = [0, 1]$, so that $c = 0.25$), where the volatility between
  both markets differed the most. Let
  \begin{equation*}
    \begin{cases}
      \theta_{X}(t) = \sigma_{X}(t) := \{\var(X_t)\}^{1/2}, \\
      \theta_{Y}(t) = \sigma_{Y}(t) := \{\var(Y_t)\}^{1/2}, \\  
    \end{cases}
  \end{equation*}
Fig.~\ref{fig:1}(c) depicts a simulated example of two artificial stock prices evolving during a period of 1 year ($T=[0,1]$), along with the 3 month period over which the volatilities in both markets differed the most in the $L^2$ sense.
\end{example}
To allow for visualizations, in the examples above we focused on low-dimensional instances but the theory to be presented next holds in general for any compact ground set $T \subset \mathbf{R}^d$. 

\subsubsection*{Preparations}
We start by laying the groundwork and recalling background.
Let
$\theta_{X} \equiv \theta_{X}(t)$ and
$\theta_{Y} \equiv \theta_{Y}(t)$ be functional
parameters characterizing $q$ different marginal features of the processes $X$ and $Y$ for $t \in T \subset \mathbf{R}^d$. Throughout, we will assume that the ground set $T$ is compact and that $\theta_{X}$ and $\theta_{Y}$ live in the Banach space $(L^p(T), \|\cdot\|_p)$, with $\|f\|_p = (\int_T |f|^p\, \dif \mu)^{1/p}$,  where $\mu$ is a measure over the Borel sets on $T$; we will refer to $\|f\|_p^{(A)} = (\int_A |f|^p\, \dif \mu)^{1/p}$ as the $L^p$ sub--norm over $A \subseteq T$.

The
learning problem from Section~\ref{problem} boils down to searching for a set $A \subseteq T$, with volume not greater than $c \geq 0$,  
over which the difference between $\theta_{X}$ and
$\theta_{Y}$ is highest. As it will be shown below, a solution to such optimization problem can be ensured to exist provided that we impose additional structure on the search domain, namely convexity and compactness. We equip the set $\mathcal{A}$ of compact and convex subsets of the ground set $T$, with the Hausdorff distance 
\begin{equation*}
  d_{\text{H}}(A, B) = \max\left\{\max_{t \in A} \,d(t, B), \max_{t \in B} \,d(t, A)\right\}, \qquad A, B \in \mathcal{A}. 
\end{equation*}
Here, $d(t, B)$ is the minimum Euclidean distance from the point $t \in A$ to the set $B$. 
Let $A \subseteq T$ be such that $|A| \leq c$, where $c \geq 0$ defines the maximum size of the regions of interest, and with $|A|$ denoting the Lebesgue measure of $A$. 


\begin{figure}
  \centering \includegraphics[scale = 0.3]{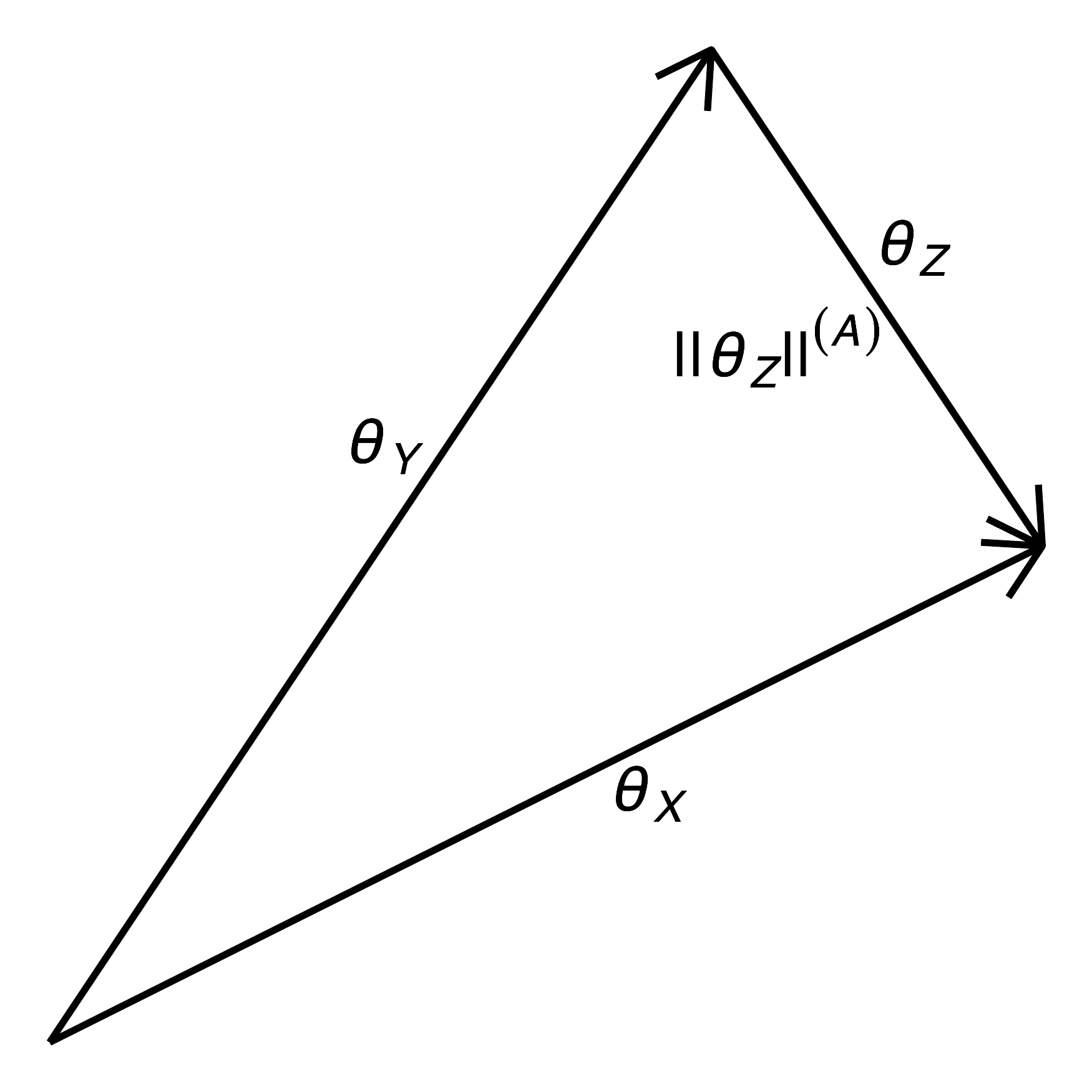}
  \caption{\label{fig:geom} \footnotesize Geometry underlying regions of maximum
    discrimination. The set function optimization problem in \eqref{eq:problem} searches 
    for the region $A$ whose $L^p$ sub-norm of $\theta_Z = \theta_{X} - \theta_{Y}$  is the largest.}
\end{figure}

\subsubsection*{Regions of maximum dissimilarity}
A compact and convex set $A \subseteq T$ is said to be a \textit{region of maximum dissimilarity, with volume bounded by $c$}, if it maximizes $F(A) = \|\theta_{X} - \theta_{Y}\|^{(A)}_p$ and if its volume does not exceed $c$; the formal definition is as follows. 

\begin{definition}[$L^p$ Region of Maximum Dissimilarity]\label{rmd}  
  Suppose $\theta_X, \theta_Y \in L^p(T)$ and $T \subset \mathbf{R}^d$ is compact. A region of maximum dissimilarity (RMD) is defined as a set $A^*_c \subseteq T$ that solves,
\begin{equation}
\label{eq:problem}
  \begin{array}{rl}
    \underset{A \subseteq T}{\max} & \|\theta_{X} - \theta_{Y}\|^{(A)}_p \\
    \textit{s.t.} & |A| \leq c, \\
    & A~\text{is compact and convex},
  \end{array}
\end{equation}
for a fixed $c \geq 0$. In addition, $D_c^*(\theta_X, \theta_Y) = \|\theta_{X} - \theta_{Y}\|^{(A^*_c)}$ is said to be the {dissimilarity index}.
\end{definition}

\noindent Some comments on Definition~\ref{rmd} are in order. Fig.~\ref{fig:geom} depicts the geometry underlying the set function optimization problem in \eqref{eq:problem}. The optimization problem in \eqref{eq:problem} is a continuous set function  optimization problem similar to \eqref{setopt}, with
\begin{equation*}
  \begin{cases}
    F(A) = \|\theta_{X} - \theta_{Y}\|^{(A)}_p, \\
    \mathscr{A} = T, \\ 
    \mathscr{F}_c = \{A: |A| \leq c, A~\text{is compact and convex}\}.    
  \end{cases}
\end{equation*}
Clearly, $F(A)$ is an increasing set function, that is if $A \subseteq B$ then $F(A) \leq F(B)$. Also, $F(A)$ is modular (or additive) in the sense that $F(A) + F(B) = F(A \cup B) + F(A \cap B)$. While it is evident that $A^*_c$ exists for a few specific examples (say, $A^*_{|T|} = T$ almost everywhere, if $T$ is convex), the existence of $A^*_c$ in general is not straightforward and it is proved below in Theorem~\ref{props}. RMDs and dissimilarity sets also have a number of attributes which we summarize over Theorem~\ref{props}. 


\begin{theorem}\label{props}
  Suppose $\theta_X, \theta_Y \in L^p(T)$ and $T \subset \mathbf{R}^d$ is compact. 
  The quantities $A_c^{*}$ and $D_c^{*}$ obey the following properties:
  \begin{enumerate}
  \item $A_c^{*}$ exists, for every $c \geq 0$.
  \item Suppose $\mu(A) = |A|$. Then, the RMDs of $(K \theta_X, K \theta_Y)$ and $(L \theta_X, L \theta_Y)$ are respectively $A^{K}_c = A^*_c$ and $A^{L}_c = A^*_{\alpha + \beta c}$, 
    where $K \theta(t) = \alpha + \beta \theta(t)$, $L \theta(t) = \theta(\alpha + \beta t)$, $\alpha \in \mathbf{R}$, and $\beta \neq 0$. 
  \item $D_c^{*}(\theta_X, \theta_Y)$ is a distance over $\Theta_c = \{\theta\restr{A}\colon \theta\in L^p(T), A \in \mathscr{F}_c\}$, for $c > 0$.
  \item $D_c^{*}(\theta_X, \theta_Y)$ is non-decreasing as a function of $c$, for fixed $\theta_X, \theta_Y \in L^p(T)$.
  \end{enumerate}
\end{theorem}

Theorem~\ref{props} warrants some comments. The existence of RMDs (Claim~1) follows from the continuity of the volume functional and the Blaschke selection theorem (Appendix~\ref{lemmata}); an RMD needs not however to be unique, as can be easily seen by considering the limiting case $\theta_{X} = \theta_{Y}$ for which every compact and convex subset of $T$ with measure $c$ is an RMD. Claim~2 shows how RMDs are impacted by a group of linear transformations acting either on the functional parameter or over time; the assumptions $\mu(A) = |A|$ and $\beta \neq 0$ are only required for $A^{L}_c = A^*_{\alpha + \beta c}$, as indeed $A_c^K = A_c^*$ holds more generally. It is proved in Claim~3 that $D_c^{*}(\theta_X, \theta_Y)$ defines a distance on the space of functional parameters in $L^p(T)$ restricted to regions in $\mathscr{F}_c$, which implies that dissimilarity indices have a metric  interpretation. Finally, Claim~4 notes that $D_c^{*} = D_c^{*}(\theta_X, \theta_Y)$ cannot decrease as $c$ increases.


\subsubsection*{Balls of maximum dissimilarity}
\noindent Let $\mathscr{B}_p$ be the family of all closed balls in $L^p(T)$, for $p \geq 1$, that is,
\begin{equation*}
  \mathscr{B}_p = \{B(t, r): r > 0, t \in T\}, 
\end{equation*}
where $B(t, r) = \{s: \|s -  t\|_p \leq r\}$; to ease notation we drop the dependence of $B(t, r)$ on $p$. A more parsimonious option for modeling is to put more structure on the shape of RMDs, and this leads us to the following definition. 
\begin{definition}[$L^p$ Ball of Maximum Dissimilarity]\label{bmd}
  Suppose $\theta_X, \theta_Y \in L^p(T)$ and $T \subset \mathbf{R}^d$ is compact. An $L^p$ ball of maximum dissimilarity (BMD) is defined as a set $B_c^* \subseteq T$ that solves,
\begin{equation}
\label{eq:problem2}
\begin{array}{rl}
  \underset{B \subseteq T}{\max} & \|\theta_{X} - \theta_{Y}\|^{(B)}_p \\
    \textit{s.t.} & |B| \leq c \\
    & B \in \mathscr{B}_p,
  \end{array}
\end{equation}
for a fixed $c \geq 0$.
\end{definition}
\noindent As in Definition~\ref{eq:problem2} we refer to $\mathscr{D}^*_c = \mathscr{D}^*_c(\theta_X, \theta_Y) = F(B^*_c)$ as the dissimilarity index for BMDs. Since the volume of an $L^p(T)$ ball of radius $r$ in $\mathbf{R}^d$ is 
\begin{equation*}
  |{B(t, r)}| = \frac{\{2 r \Gamma(1 / p + 1)\}^d}{\Gamma(d / p + 1)},
\end{equation*}
for all $t \in T$, where $\Gamma(z) = \int^\infty x^{z - 1} \text{e}^{-x}\, \dif x$ is the Gamma function, it follows that the volume constraint on BMDs, $|B| \leq c$, can be rewritten as a function of the radius, that is
\begin{equation}
  \label{rstar}
  r \leq \frac{\{c \Gamma(d / p + 1)\}^{1 / d}}{2 \Gamma(1 / p + 1)} \equiv R_c.
\end{equation}
Given the similarities between the definitions of RMDs and BMDs it is not surprising that have identical properties
(say, for $c > 0$, $\mathscr{D}^*_c(\theta_X, \theta_Y)$ is a also distance 
and $\mathscr{D}_c^{*}$ is non-decreasing). The line of attack
for proving existence of BMDs is analogous to the one
from Theorem~\ref{props} for RMDs but at this time
Tikhonov's theorem (Appendix~\ref{lemmata}) implies 
compactness of the search domain. And interestingly the smoothness of $R_c$ is inherited by $\mathscr{D}_c^{*}$.
\begin{theorem}\label{Dc2}
  Suppose $\theta_X, \theta_Y \in L^p(T)$ and $T \subset \mathbf{R}^d$ is compact. 
  The quantities $B_c^{*}$ and $\mathscr{D}_c^{*}$ obey the following properties:
  \begin{enumerate}
  \item $B^*_c$ exists, for every $c \geq 0$.
  \item $\mathscr{D}_c^{*}$ is continuous, and the ``argmax'' correspondence of center--radius, $\alpha_c: [0, \infty) \twoheadrightarrow T \times [0, R_c]$, defined as 
      $\alpha_c = \{(t, r) \in T \times [0, R_c]: f(t, r) = \mathscr{D}_c^{*}\}$,
    is upper hemicontinuous for every $c \geq 0$, where $f(t, r) = F\{B(t, r)\}$.
  \end{enumerate}
\end{theorem}

\subsubsection*{A modicum on computing and numerical optimization}
An important consequence of \eqref{rstar} is that the set function constrained optimization problem \eqref{eq:problem2} that leads to BMDs, can actually be written as a standard continuous optimization problem over $T \times [0, R_c] \subset \mathbf{R}^{d + 1}$. Indeed, it follows from \eqref{rstar} that \eqref{eq:problem2} is equivalent to computing
\begin{equation}
\label{maxb}
  {\max}\{f(t, r): (t, r) \in T \times [0, R_c]\}. \\
\end{equation}
where $f(t, r) = F\{B(t, r)\}$. That is, the BMD is ${B(t_{c}^*, r^*_c)}$ with $(t_{c}^*, r^*_c)$ maximizing \eqref{maxb}, and hence in practice BMDs can be computed via derivative free optimization algorithms such as conjugate search, implicit filtering, pattern search, or Nelder--Mead \citep[][Ch.~9]{nocedal2006}. When the center of the optimal BMD ($t^*_{c}$) is `sufficiently far' from the boundary of $T$, then the optimal radius ($r^*_c$) is $R_c$ in \eqref{rstar}. That is, when $d(t^*_{c}, \partial T) > R_c$ the optimal radius is $r^*_c = R_c$ (as $f(t, r)$ is a non-decreasing function of $r$) in which case the optimization problem in \eqref{maxb} resumes to searching for $t^*_{c} \in \mathbf{R}^d$. This also implies that often in practice the marginal posterior of $r^*_c$ is essentially degenerated. 

\subsection{Learning from Data}\label{learningfromdata}
\subsubsection*{Latent Gaussian model specification}
To model BMDs in applications we consider a version of the latent Gaussian model specification in \cite{rue2009} adapted to our setup; to ease notation, below we only refer to $X(t)$, and denote its functional parameter by $\theta(t) \equiv \theta_X(t)$, but all comments apply to $Y(t)$ as well. A latent Gaussian model is essentially a Bayesian generalized additive model that assigns Gaussian priors to  parameters and a possibly non-Gaussian prior to its hyperparameters. Specifically, suppose that $Z(t) = h\{X(t)\}$ is in the exponential family, with its mean function coinciding with the functional parameter, and that 
\begin{equation}\label{gspec}
  \theta(t) = g\bigg(\beta_0 + \sum_{i=1}^B\beta_i\phi_i(t)\bigg). 
\end{equation}
Here $\{\phi_i \equiv \phi_i(t)\}_{i= 1}^B$ is a set of basis functions in $L^p(T)$, $\beta = (\beta_0, \dots, \beta_B)^{\T}$ is a parameter, and $g$ is an inverse link function.~Following \citeauthor{rue2009}~we assign a multivariate Normal prior with a sparse precision matrix ($Q$) to $\beta$, 
 which induces a Gaussian process prior on $g^{-1}(\theta(t))$ with a conditional independence property. Many functional parameters can be modeled in this way including those from Examples~\ref{ex:1}--\ref{ex:3} and all numerical instances from Sections~\ref{simulation}--\ref{data}. The theoretical developments from Section~\ref{sec:dis} apply however more generally beyond the modeling assumptions made over this section. 

 \subsubsection*{Inla-based inference for balls of maximum dissimilarity}
 We now discuss how to conduct inference for balls of maximum dissimilarity. It is well-known that the latent Gaussian model described above can be fitted with an Integrated Nested Laplace Approximation (INLA) \citep{rue2009}; the method is effective even when the dimension of the precision matrix $Q$ is large, and is particularly tailored for the case where the number of hyperparameters, $\alpha$, is moderate (say, 6--12). INLA is a deterministic method for approximating the marginal posterior of each parameter that is based on the Laplace approximation; loosely speaking, the Laplace approximation is an approximation for integrals of the type $\int e^{-n f(y)}\, \dif y$ for large $n$, that approximates the integrand ($e^{-n f(y)}$) with a Gaussian density centered at its mode and sets the covariance matrix as the inverse of the curvature (around the mode); see, for instance, \citet[][Section~9.7]{young2005}. Below, we sketch some brief details on INLA; further details can be found elsewhere \citep{rue2009, blangiardo2015, rue2017, wang2018, krainski2018, gomez2020}. The first step of INLA approximates the marginal posterior of $\alpha$ via the Laplace approximation, that is, 
\begin{equation}\label{step1}
  \begin{split}
    p(\alpha \mid \text{data}) &= \frac{p(\alpha, \beta \mid \text{data})}{p(\beta \mid \alpha, \text{data})} \\
    &\approx \frac{p(\alpha, \beta \mid \text{data})}{\widetilde{p}(\beta \mid \alpha, \text{data})}\bigg|_{\beta = \beta_\alpha^*},     
  \end{split}
\end{equation}
where $\widetilde{p}(\beta \mid \alpha, \text{data})$ is the Gaussian approximation based on the mode---and the curvature around the mode---of the full conditional of $\beta$, and where $\beta_\alpha$ is the mode of this approximated full conditional for a given $\alpha$. Next, INLA approximates the marginal posterior of each component of $\beta$. Let $\beta_{-i}$ be the elements of $\beta$, except $\beta_i$. Similarly to \eqref{step1} it follows that
\begin{equation}\label{step2}
  p(\beta_i \mid \alpha, \text{data}) \propto \frac{p(\alpha, \beta \mid \text{data})}{p(\beta_{-i} \mid \alpha, \beta_i, \text{data})},
\end{equation}
which can be approximated using a Laplace approximation for $p(\beta_{-i} \mid \alpha, \beta_i, \text{data})$; faster approximations are also available from \cite{rue2009}. Finally, the marginal posterior density $p(\beta_i \mid \text{data})$ is obtained by numerically integrating out $\alpha$. Independent samples from the full posterior of $\betab$ can then be generated following \citet[][Section~2.5.2]{seppa2019}, which can then be used for estimating functionals of the parameters of interest.

Estimation and inference for balls of maximum dissimilarity can be conducted with an algorithm that combines the deterministic nature of INLA along with sampling, according to the steps below. Data from processes $X$ and $Y$ are respectively denoted by $\mathcal{D}_X$ and $\mathcal{D}_X$. 

\begin{algorithm}
\caption{INLA-Based Posterior Inference for BMDs\label{alg}}   
\begin{algorithmic}
\item[~~~Step~1:] Fit the marginal posterior densities, $p(\beta_{X, i}\,|\,\mathcal{D}_X)$ and $p(\beta_{Y, i}\,|\,\mathcal{D}_Y)$, using the Integrated Nested Laplace Approximation, for $i = 0, \dots, B_X$ and $j = 0, \dots, B_Y$. 
\item[~~~Step~2:] Sample $m$ posterior draws from the full posterior of $\beta_X = (\beta_{X, 0}, \dots, \beta_{X, B_X})$ and $\beta_Y = (\beta_{Y, 0}, \dots, \beta_{Y, B_Y})$, given by 
  $$\beta_{X}^{(1)},\dots,\beta_{X}^{(m)} \, \iid \, p(\beta_{X}\,|\,\mathcal{D}_{X}),
  \qquad \beta_{Y}^{(1)},\dots,\beta_{Y}^{(m)} \,\iid\,  p(\beta_{Y}\,|\,\mathcal{D}_Y),$$
  so to generate $m$ posterior trajectories from the functional parameters using \eqref{gspec}; that is, for $k = 1, \dots, m$, do 
  \begin{equation}\label{trajs}
    \theta_X^{(k)}(t) = g\bigg(\beta_{X, 0}^{(k)} + \sum_{i=1}^{B_X}\beta_{X, i}^{(k)}\phi_i(t)\bigg), \qquad 
    \theta_Y^{(k)}(t) = g\bigg(\beta_{Y, 0}^{(k)} + \sum_{j=1}^{B_Y}\beta_{Y, j}^{(k)}\phi_j(t)\bigg). 
  \end{equation}
\item[~~~Step~3:] Use the posterior trajectories from Step~2 to obtain a sequence of posterior BMD draws for $$\{B_c^{(k)*} \equiv   {B(t_{c}^{(k)*}, r_c^{(k)*})}\}_{k=1}^m,$$ with $(t_{c}^{(k)*}, r_c^{(k)*})$ solving the optimization problem in \eqref{maxb}, for $k = 1, \dots, m$.
\end{algorithmic}
\end{algorithm}
Algorithm~\ref{alg} warrants some comments. Step~1 is deterministic, it follows by numerically integrating out the hyperparameters in \eqref{step2}, and to facilitate its implementation we recommend using the \texttt{R-INLA}~package \citep{martins2013, lindgren2015}  from \texttt{R} \citep{rdevelopmentcoreteam2016}, which is also equipped with routines that facilitate the implementation of Step~2 (e.g.~\texttt{inla.posterior.sample}). Step~3 boils down to solving the optimization problem in \eqref{maxb} using the pair of posterior trajectories in \eqref{trajs}. Except where mentioned otherwise, in all experiments reported below we draw $m = 1\,000$ times from the posterior distribution of BMDs using Algorithm~\ref{alg}. 

\section{Variants, Consequences, and Extensions}\label{extensions}

\subsection{Multi-Objective RMDs}\label{multi}
We now consider the context where the interest is on learning about  
regions over which a set of marginal features of two processes most
differ. Let
$$
\begin{cases}
\theta_{X} = (\theta_{X, 1}, \dots, \theta_{X, q}) = (\theta_{X, 1}(t), \dots, \theta_{X, q}(t)),  \\
\theta_{Y} = (\theta_{Y, 1}, \dots, \theta_{Y, q}) = (\theta_{Y, 1}(t), \dots, \theta_{Y, q}(t)),  
\end{cases}
$$ be functional
parameters characterizing marginal features of these processes.  The
learning problem consists of searching for a set $A$, with volume $|A|\leq c$, over which the overall difference between $\theta_{X}$ and
$\theta_{Y}$ is highest. Formally, the concept below entails the combination of two fields of optimization on whose interface little is known, namely: Multi-objective optimization \citep[e.g.][]{pardalos2017non} and set function optimization. In this section
\begin{equation}
  \label{eq:fi}
  F_i(A) = \|\theta_{X, i} - \theta_{Y, i}\|^{(A)},
\end{equation}
denotes the set objective functions of interest, for $i = 1, \dots, q$.

\begin{definition}[Region of Multi-Maximum Dissimilarity]\label{rmmd}
  Suppose $\theta_X, \theta_Y \in L^p(T)$ and $T \subset \mathbf{R}^d$ is compact. 
The region of multi-maximum dissimilarity (RMMD) is defined as a set $A^*_c$ that solves,
\begin{equation}
\label{eq:multitar}
  \begin{array}{rl}
    \arg \underset{A \subseteq T}{\max} &(F_1(A), \dots, F_q(A))\\
    \textit{s.t.} & |A| \leq c, \\
                  & A~\text{is compact and convex},
  \end{array}
\end{equation}
for a fixed $c \geq 0$, where $F_i(A)$ is defined as in \eqref{eq:fi}. In addition,  $D_c^* = (F_1(A^*_c), \dots,F_q(A^*_c))$ is said to be the multi-dissimilarity index.
\end{definition}Ideally, we would aim to simultaneously maximize all targets $F_i(A) = \|\theta_{X, i} - \theta_{Y, i}\|^{(A)}$, for all $i$. Yet, in practice targets may conflict each other, that is, if one is increased some other may be decreased; we illustrate this situation in Section~\ref{sec:vol}. To give a more concrete meaning to Definition~\ref{rmmd}, an ordering concept over the collection of feasible sets $\mathscr{F}$ is required, so to rank the corresponding objective function values. The following Pareto optimal concept induces that order, and it defines optimality via a compromise across all objective set functions, in the sense that improvements on one target cannot be made at the cost of deteriorating another target. 
\begin{definition}[Pareto Optimal Region of Multi-Maximum Dissimilarity] Let $F_i(A)$ be defined as in \eqref{eq:fi}. The set $A^*_c \in \mathscr{F}_c$ is a Pareto optimal region of multi-maximum dissimilarity if there exists no other set $A \in \mathscr{F}_c$ such that  $F_i(A) \geq F_i(A^*_c)$, for all $i \in \{1,\dots, q\}$, and $F_i(A) > F_j(A^*_c)$, for at least one $j \in \{1,\dots, q\}$.
  \end{definition}

In common with standard theory for multi-objective function, in our set function context the set of Pareto optimal RMMDs can be analytically obtained only in very specific cases, and thus we need to resort to scalarization  \citep[][Ch.~2]{pardalos2017non}. The aim of scalarization is to reduce a multi-objective problem into a single-objective problem. Here, we define and characterize the following linear scalarization method for our set function context 
\begin{equation}\label{Fw}
\mathcal{F}_{w}(A)=\sum_{i=1}^q w_i F_i(A), 
\end{equation}
$w=(w_1,\dots,w_q) \in (0, \infty)^q$, and we will refer to 
\begin{equation}\label{soll}
  A_{w,c}^*  = \arg \underset{A \in \mathscr{F}_c}{\max}\, \mathcal{F}_{w}(A),
\end{equation}
as the \textit{solution to the set function linear scalarization problem} with weight $w$. 
\begin{theorem}\label{scal}
  Every solution to the linear scalarization problem is a Pareto optimal RMMD.
\end{theorem}
Section~\ref{sec:vol} illustrates how Theorem~\ref{scal} can be applied in practice. 

\subsection{RMDs-Based on Averaging}\label{densities}
This section shows that a simple modification of the notion of RMDs based on `averaging'
leads to links with other well-known concepts.
To streamline the presentation of ideas we will focus in $L^1(T)$, and thus exceptionally over this section we will write $B(t, r)$ to denote $\{t: \|t -  t\|_1 \leq r\}$, and we will further set $\mu(A) = |A|$. Averaging is here understood in the usual sense of the well-known average integral symbol, $\dashint$, which is defined as
\begin{equation}\label{aveint}
  \dashint_{B(t, r)} f(u) \, \dif u  =
  \frac{1}{|B(t, r)|}\int_{B(t, r)} f(u) \, \dif u, 
\end{equation}
for an absolutely integrable $f$ and for $t \in T$ and $r \geq 0$. For $r = 0$, the expression in \eqref{aveint} should be understood as the limit,  $\dashint_{B(t, r)} f(u) \, \dif u \to f$, as $r \to 0$. There are two motifs in this section: 
\begin{itemize}
\item \textit{Averaging the target}: In the optimization problem that defines BMDs (Definition~\ref{bmd} and Equation~\eqref{maxb}), we now replace $\int$ with $\dashint$ in the objective function.
\item \textit{Moving beyond random processes}: While the focus in the previous sections has been on modeling RMDs and BMDs for time and spatial processes, it would seem natural to
employ these concepts beyond the context of stochastic
processes. Thus, in this section $\theta_{X}(t)$ will $\theta_{Y}(t)$
will be allowed to be general parameters, possibly unrelated with stochastic
processes, such as distribution functions associated with 
random variables $X$ and $Y$.
\end{itemize}
We define the \textit{averaged} or \textit{Hardy--Littlewood BMD} as a ball, $\mathbb{B}_c^* = B(t_{c}^*, r^*_c)$, that maximizes, 
\begin{equation}\label{HLd}
  \mathbb{D}_c^*(\theta_X, \theta_Y) =
  \max\bigg\{\dashint_{B(t, r)} |\theta_X(u) - \theta_Y(u)| \, \dif u 
  : (t, r) \in T \times [0, R_c]\bigg\},
\end{equation}
where $\theta_X, \theta_Y$ are absolutely integrable and $c \geq 0$. As can be seen by comparing \eqref{HLd} with \eqref{maxb} the only difference between Hardy--Littlewood BMDs and standard BMDs in \eqref{maxb} is the use of the average integral symbol in \eqref{HLd} rather than the standard integral $\int$.\footnote{We will refer to these BMDs as ``Hardy--Littlewood'', given that $\mathbb{D}_c^*(\theta_X, \theta_Y)$ in \eqref{HLd} has links with the so-called Hardy--Littlewood maximal function \citep[e.g.][Section~1.6]{tao2011},
\begin{equation*}
(Mg)(t) = \sup \bigg\{\dashint_{B(t, r)} |g(u)| \, \dif u: r > 0 \bigg\},  
\end{equation*}
where $g$ is absolutely integrable.} The fact that $\mathbb{B}_c^*$ exists follows by noticing that $f(t, r) = \dashint_{B(t, r)} |\theta_X(u) - \theta_Y(u)| \, \dif u$ is continuous and that Tikhonov's theorem (Appendix~\ref{lemmata}) implies that the search domain, $T \times [0, R_c]$, is compact for every $c \geq 0$, as both $T$ and $[0, R_c]$ are themselves compact. In addition, the following proposition holds. 
\begin{proposition}\label{dhlprop}
  Suppose $\theta_X, \theta_Y \in L^1(T)$ are continuous and $T \subset \mathbf{R}^d$ is compact. Then, for every $c > 0$ it holds that 
  $\mathbb{D}_c^*(\theta_X, \theta_Y) \leq \max_{t \in T} |\theta_X(t) - \theta_Y(t)|$, and as $c \to 0$, $$\mathbb{D}_c^*(\theta_X, \theta_Y) \to \underset{t \in T}{\max} \,|\theta_X(t) - \theta_Y(t)|.$$ 
\end{proposition}
\noindent
The next example uses the lenses of Proposition~\ref{dhlprop} to show that a well-known 
  performance measure known as Youden index  
  \citep[e.g.][]{sokolova2006, inaciodecarvalho2017} has links with the averaged BMDs defined in \eqref{HLd}.
\begin{example}[Youden index and Kolmogorov metric]\normalfont \label{youex}  
   If
  $\theta_{X}(t) = F_{X}(t)$ and
  $\theta_{Y}(t) = F_{Y}(t)$ are distribution
  functions supported on $T \subset \mathbf{R}$, and if we set $c \to 0$, then
  maximization takes place over singletons and the 
  Lebesgue differentiation theorem (Appendix~\ref{lemmata}) 
  yields that 
  \begin{equation*}
    \lim_{r \to 0} \, 
    \dashint_{B(t, r)} |F_X(u) - F_Y(u)| \, \dif u = |F_{X}(t) - F_{Y}(t)|. 
\end{equation*}
  Thus, when $c \to 0$ the set function
  optimization problem in \eqref{HLd}, 
  becomes a standard continuous optimization problem that can be written as
\begin{equation}\label{YI}
  \max_{t \in T} |F_{X}(t) - F_{Y}(t)|, 
\end{equation}
which is the well-known Youden index. Thus, the Youden index is an Hardy--Littlewood dissimilarity
index, and the popular Youden's $\text{J}$ statistic (i.e. the maximizer of \eqref{YI}) is a singleton
averaged BMD, as in \eqref{HLd} with $c \to 0$. Another consequence of \eqref{YI} is that the limiting
Hardy--Littlewood distance in \eqref{HLd} when $c \to 0$  
is the well-known Kolmogorov metric 
\citep[e.g.,][]{gretton2012kernel}.
\end{example}

\section{Numerical Study with Artificial Data}\label{simulation}
In this section we assess the performance of the proposed tools via a Monte Carlo study.

\subsubsection*{Artificial data generating processes and simulation settings}
Examples~\ref{ex:1}--\ref{ex:2} from Section~\ref{sec:dis} will form the basis of this numerical workout. Namely, we consider the following scenarios. \\

\noindent \textbf{Scenario~1 (Mean Functions)} BMDs between mean functions as in Example~\ref{ex:1}. Here, $X_t$ and $Y_t$ are Gaussian processes  
  with mean functions given in \eqref{funex1}   
  and where the same Mat\'ern covariance function is assumed for both processes. Specifically
  $\text{cov}(X_s,X_t) = \text{cov}(Y_s, Y_t) = C_\nu(\|s-t\|_2, \sigma, \ell)$, where 
  \begin{equation*}
    \begin{split}
      C_\nu(d, \sigma, \ell) = 
       \sigma^2 \frac{2^{1-\nu}}{ \Gamma(\nu)}\bigg(\sqrt{2\nu}\frac{d}{\ell}\bigg)^\nu K_\nu\bigg(\sqrt{2\nu}\frac{d}{\ell}\bigg),
    \end{split}
  \end{equation*}
  for $(s, t) \in [0, 1]^2$, where $K_\nu$ is the modified Bessel function \cite[][Section~9.6]{abramowitz1964}, and where $\sigma, \nu$, and $\ell$ are positive parameters, here set as $\sigma = \nu = \ell = 1$. The simulated data are then a discretized version of $n$ simulated Gaussian processes evaluated over a grid on the unit interval, 
  $$\mathcal{D}_X = \{X_{t, i}: t \in \{0/J, \dots, (J - 1)/J\}\}_{i = 1}^n,$$
  with  $n\in \{10,50,100,200\}$ and $J \in \{10, 20\}$; the same comments apply to $\mathcal{D}_Y$. \\

\noindent \textbf{Scenario~2 (Intensity Functions)} BMDs between intensity functions as in Example~\ref{ex:2}. Here, points drawn from non-homogeneous bivariate Poisson process with mean measures, 
\begin{equation}\label{Lambda}
  \begin{cases}
    E\{N_X(A)\} = \int_A \gamma\exp\{ -(t_1^2 +t_2^2)/2  \} \, \dif t,  \\
    E\{N_Y(A)\} = \delta E\{N_X(A)\}, 
  \end{cases}    
\end{equation}
for $A \subseteq T = [-3, 3]^2$. 
While the sample sizes in this scenario are random quantities, given by $N_X = N_X(T)$ and $N_Y = N_Y(T)$, the mean number of simulated points over $T$ is $(E(N_X), E(N_Y)) \approx (2\pi\gamma, 2\pi\gamma\delta)$, a simple yet accurate approximation that follows immediately from multiple Gaussian integrals. The simulated data are given by the following collection of points
\begin{equation*}
  \mathcal{D}_X = \{(X_{1, 1}, X_{2, 1}), \dots, (X_{1, N_X}, X_{2, N_X})\},
\end{equation*}
with $\gamma \in \{25, 50\}$ and $\delta \in \{2, 4, 8, 16\}$, and where $\mathcal{D}_Y$ is analogously defined. \\

\subsubsection*{Modeling, prior specification, and posterior inference}
Inferences for the BMDs were carried out by sampling $m = 1\,000$ and 500 times using Algorithm~\ref{alg} for Scenarios~1 and 2, respectively. As can be seen from Algorithm~\ref{alg}, inferences for BMDs are constructed from the functional parameters, and thus we now comment on what versions of  \eqref{gspec} have been used for fitting the latter. For Scenario~1 the identity link function was used in \eqref{gspec} along with B-spline basis, and the number of basis functions was selected using the DIC (Deviance Information Criterion; \citealt{spiegelhalter2002, spiegelhalter2014}). The default uninformative priors of \texttt{R-INLA} have been used, which consist of diffuse priors for the $\beta$'s---i.e.~$\beta_0 \sim N(0, \infty)$ and $\beta_i \sim N(0, 1000)$---and a long-tailed prior for the variance of the error term---i.e.~a log gamma distribution, where the gamma distribution has mean $a/b$ and variance $a / b^2$, with $a = 1$ and $b = 10^{-5}$; see \citet[][Section~5.2.1]{wang2018} for further details. For Scenario~2 we follow \cite{simpson2016} and specify a log-Gaussian Cox process using \eqref{gspec} by setting a log link function, that links the intensity function with a Mat\'ern random field using piecewise linear basis functions over a mesh, 
and where the $\beta$'s are Gaussian-distributed. For the parameters of the Mat\'ern covariance function we use the PC prior approach of \citet{fuglstad2019} setting $P(\sigma > 1) = 0.001$ and $P(\ell < 0.05) = 0.001$. 

Before we move to the Monte Carlo study, we first illustrate the methods on a single run experiment for some instances of Scenarios~1 and 2. 

\begin{figure}[H]
  \centering \textbf{Scenario~1}\\
  \begin{tabular}{cc}
    \subfloat[]
    {\includegraphics[scale=0.19]{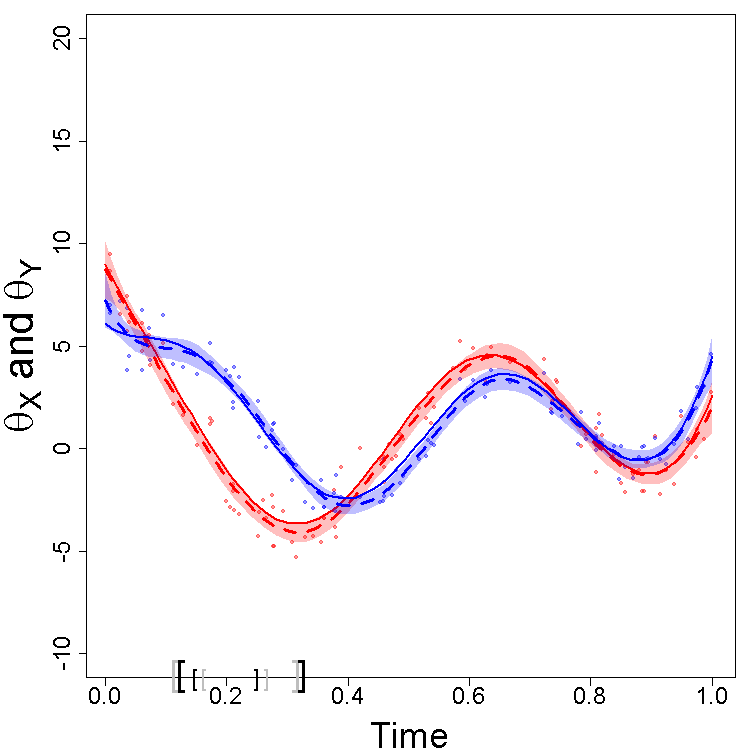}}
    \subfloat[]
    {\includegraphics[scale=0.19]{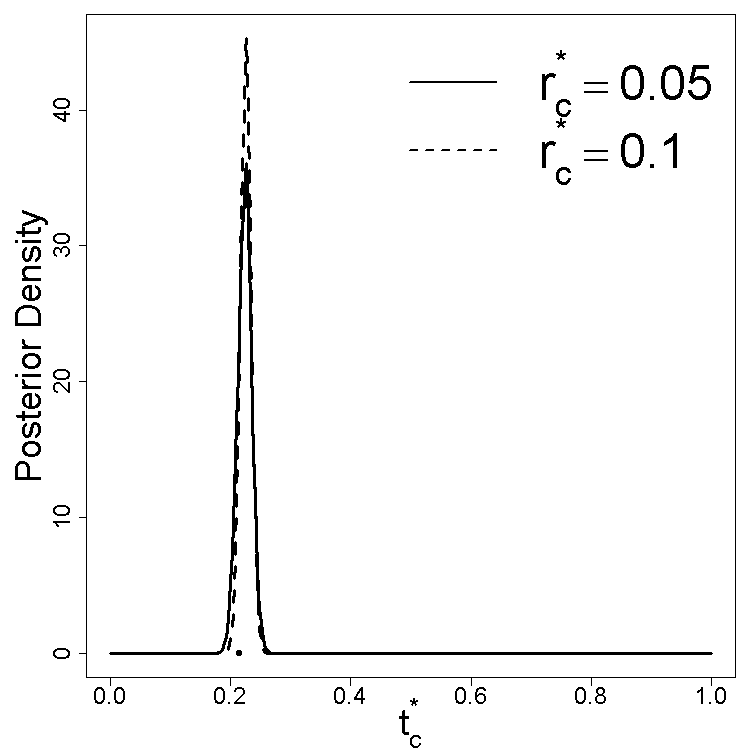}}
  \end{tabular}
  \centering \\
  \centering \textbf{Scenario~2}\\
  \begin{tabular}{cc}
    \subfloat[]
    {\includegraphics[scale=0.19]{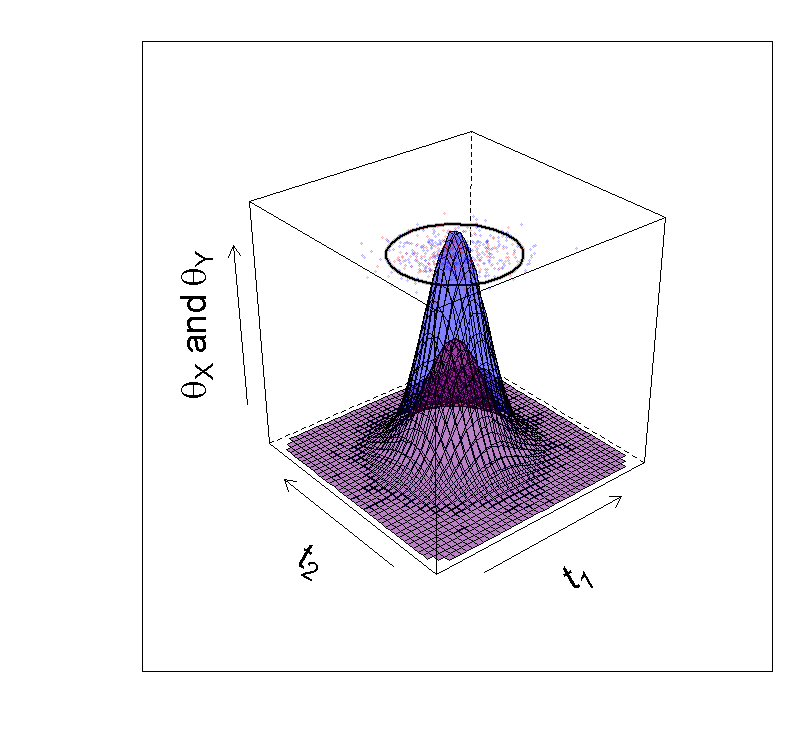}} \hspace{0.19cm}
    \subfloat[]
    {\includegraphics[scale=0.19]{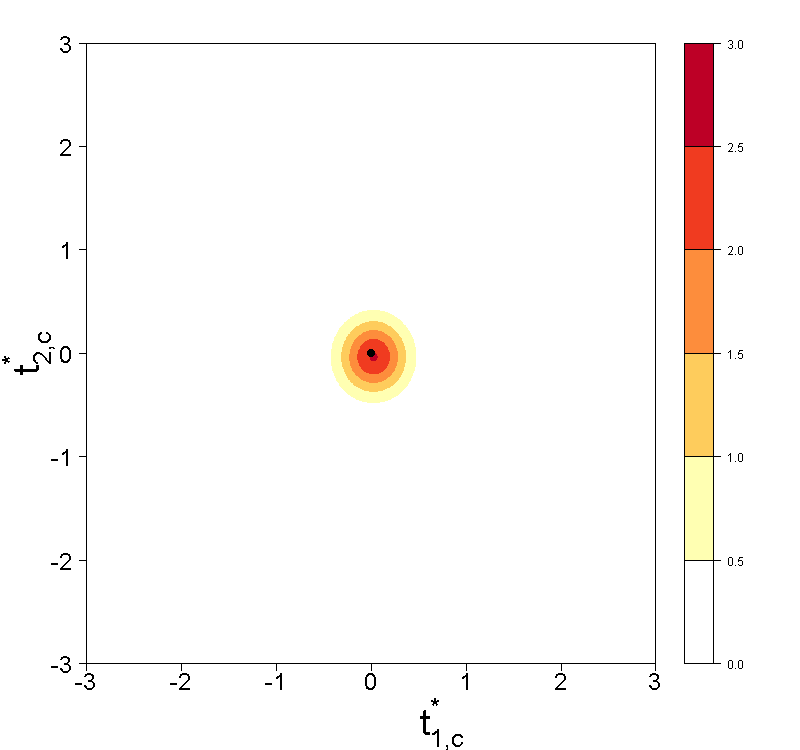}}
\end{tabular}
\caption{One shot experiments for Scenarios~1 and 2 (a) The fitted BMDs ([black]) based on the simulated Gaussian process data are depicted in the time axis and compared against the true (\gray{[gray]}); the chart also depicts the fitted functional parameters (dashed) along with $95\%$ credible bands, and the true parameters (solid). (b) Marginal posterior density for the center $t^*_{c}$ plotted against the true center (rug). (c) and (d) are identical to (a) and (b), respectively, but for the simulated point process data from Scenario~2. \label{fig:2}}
\end{figure}
 
\subsubsection*{One shot experiments}
Let's start with Scenario~1 and consider for illustration BMDs with length $c = 0.1$ and $c = 0.2$. In Fig.~\ref{fig:2}(a) we show the fitted BMDs, along with the corresponding mean functions, on a one shot experiment with $n=10$ and $J=10$. As can can be seen from Fig.~\ref{fig:2}(a), the fitted BMDs accurately recover the true $B^{*}_{0.1} = [0.165, 0.265]$ and $B^{*}_{0.2} = [0.115, 0.315]$. In Fig.~\ref{fig:2}(b) we also display the marginal posterior density for the optimal center $t^*_{c}$ which quantifies the uncertainty surrounding the true. The marginal posterior for the radius is essentially degenerated, as predicted earlier in the comments surrounding \eqref{maxb}, and hence not shown.


Let's now move to Scenario~2 and consider for illustration a BMD with area $c = 2 \pi$. In Fig.~\ref{fig:2}(c) we depict the fitted BMD, on a one shot experiment with 
$\gamma=25$ and $\delta = 2$, and display the fitted intensity functions. As it is evident from Fig.~\ref{fig:2}(c), the fitted BMD nicely uncovers the true, and indeed it completely overlaps it to the point that the true (depicted in gray) is barely visible. 

\subsubsection*{Monte carlo evidence}
We now report the main findings of the Monte Carlo simulation study. We redo the previous one shot analyses $M =1\,000$ times, considering different samples sizes, and relying on the GHE (Posterior Mean \textbf{G}lobal \textbf{H}ausdorff \textbf{E}rror), 
\begin{equation}
  \label{PMGHE}
  \text{GHE} = E\bigg\{\int^{|T|}D_{\mathrm {H} }(B^*_c,\widehat{B}^*_c) \,\dif c \, \bigg\vert \, \mathcal{D}_X, \mathcal{D}_Y\bigg\}
\end{equation}
so to quantify how accurate on average 
are the estimated BMDs, $\widehat{B}^*_c$, over $0 \leq c \leq |T|$.

\begin{figure}[H]
  \centering
  \begin{tabular}{cc}
    \textbf{Scenario~1}   \hspace{4.3cm}  \textbf{Scenario~2} \\
    \includegraphics[scale=0.35]{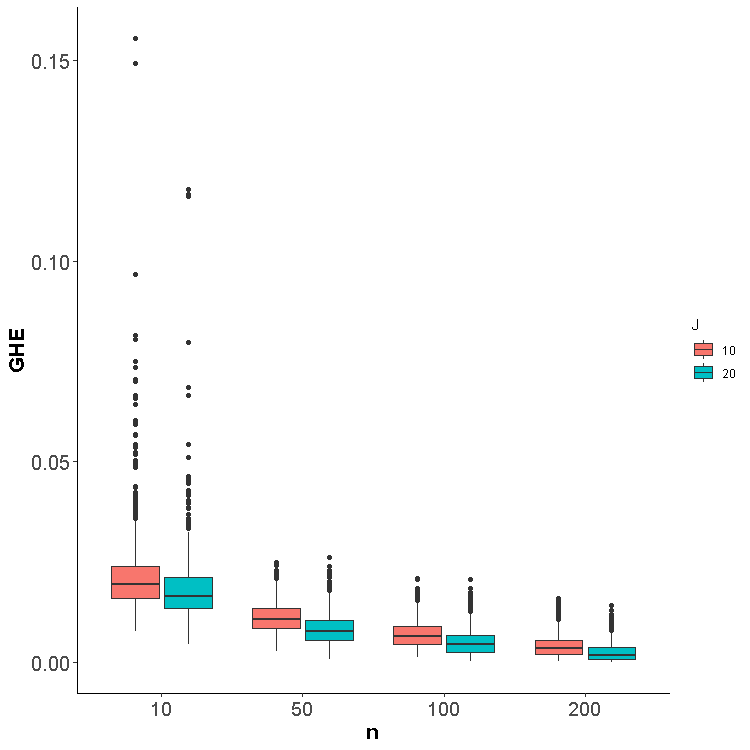}
    \includegraphics[scale=0.35]{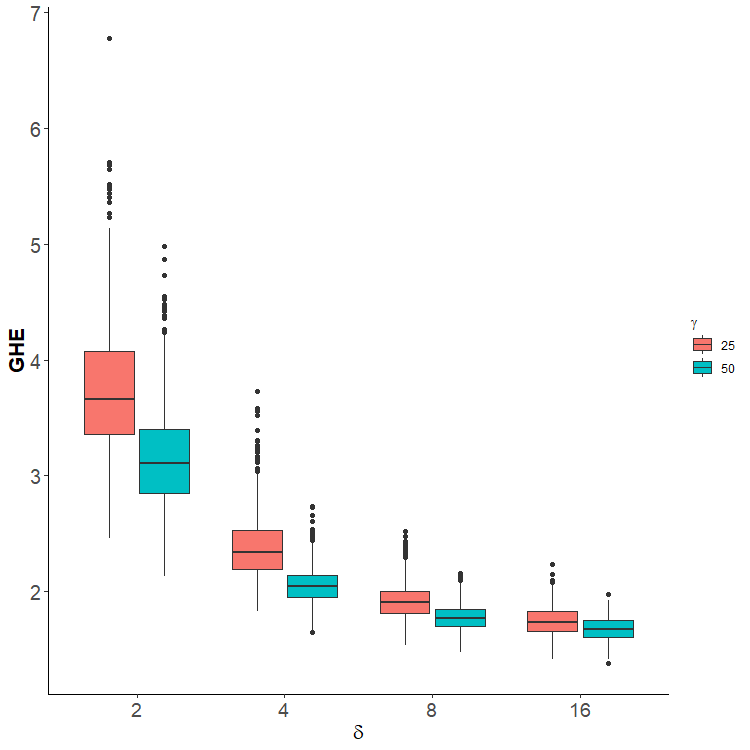}
\end{tabular}
\caption{Side-by-side boxplots of GHE (Posterior Mean \textbf{G}lobal \textbf{H}ausdorff \textbf{E}rror) for Monte Carlo simulation study.\label{mcresults}}
\end{figure}

Some comments on the computation of \eqref{PMGHE} are in order. 
In Scenario~1 we use $D_{\mathrm {H} }(B^*_c ,\widehat{B}^*_c)=\max \{|(t_{0,c}^* - r^*_c)-(\widehat{t}_{0,c}^* - \widehat{r}^*_c)|,|(t_{0,c}^* + r^*_c)-(\widehat{t}_{0,c}^* + \widehat{r}^*_c)|\}$, while in Scenario~2 we use a numerical approximation of $D_{\mathrm {H} }(B^*_c ,\widehat{B}^*_c)$ implemented using \cite{pracma}. 
Finally, the GHE for each simulated dataset is computed as 
$$\text{GHE}_j = \frac{1}{m} \sum_{i=1}^m\int^{|T|}D_{\mathrm {H} }(B^*_c,\widehat{B}_{c}^{*[i,j]})\,\dif c,$$
where $\widehat{B}_{c}^{*[i,j]}$ is the $i$th posterior sampled BMD based on the $j$th simulated sample, for $i=1,\dots,m$ and $j=1,\dots,M$. As can be seen from Fig.~\ref{mcresults}, GHE tends to decrease as $n, J, \gamma$, and $\delta$ increases. Such behavior confirms the expected frequentist behavior of the methods, as $n$ and $J$ dictates the amount of simulated data for Scenario~1, and $\gamma$ and $\delta$ does the same for Scenario~2. To put it differently, since larger values of these parameters imply larger sample sizes, the observed reduction in GHE as a function of the latter parameters suggests a sensible asymptotic performance of the proposed Bayesian inferences.

\section{Empirical Section}\label{data}


\subsection{Thefts in Buenos Aires}\label{sec:bsas}
Buenos Aires is the most dense metropolis in Argentina and its crime rates are significantly higher in comparison to the rest of the country. In this section we illustrate how the proposed method can reveal regions of the city where nonviolent crimes---such as burglary, pickpocketing or nonviolent thefts---have changed the most, comparing the years 2019 (pre COVID-19) and 2020 (when the COVID lockdown took place during several months). The data are publicly available online in the \href{https://data.buenosaires.gob.ar/dataset/delitos}{city hall web page}, and consist of point process data on the latitude and longitude where thefts occurred during 2019 ($\mathcal{D}_{2019}$) and 2020 ($\mathcal{D}_{2020}$). Here, the functional parameters of interest are the intensity functions
$$\theta_{2019}(\texttt{latitude}, \texttt{longitude}), \quad  \theta_{2020}(\texttt{latitude}, \texttt{longitude}),$$
and its BMD will represent region of the city, of a given size, where the most noteworthy changes in thefts took place. The fitted BMDs were modeled according to \eqref{gspec} using again a log-Gaussian Cox process and a similar PC prior specification as in  Section~\ref{simulation}.

\begin{figure}[h]
  \centering
  \subfloat[]
  {\includegraphics[scale=0.3]{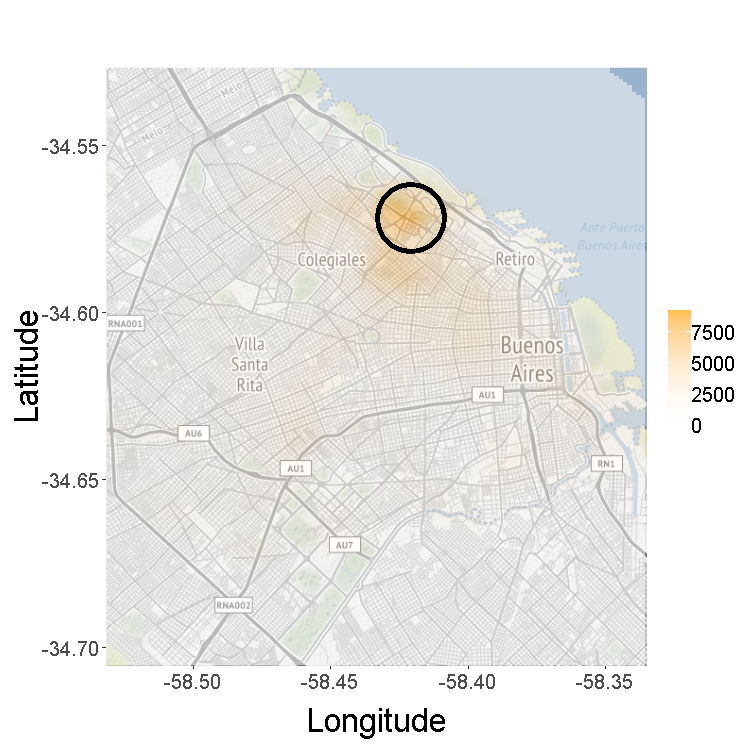}
  } 
  \subfloat[]
  {\includegraphics[scale=0.3]{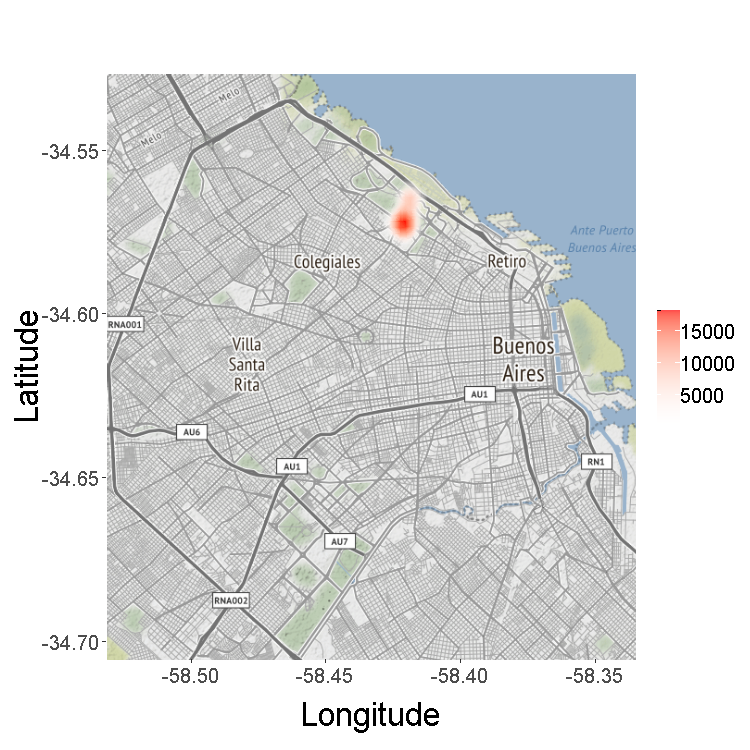}
  } \\  
  \begin{tabular}{cc}
    \subfloat[]
    {\includegraphics[scale=0.3]{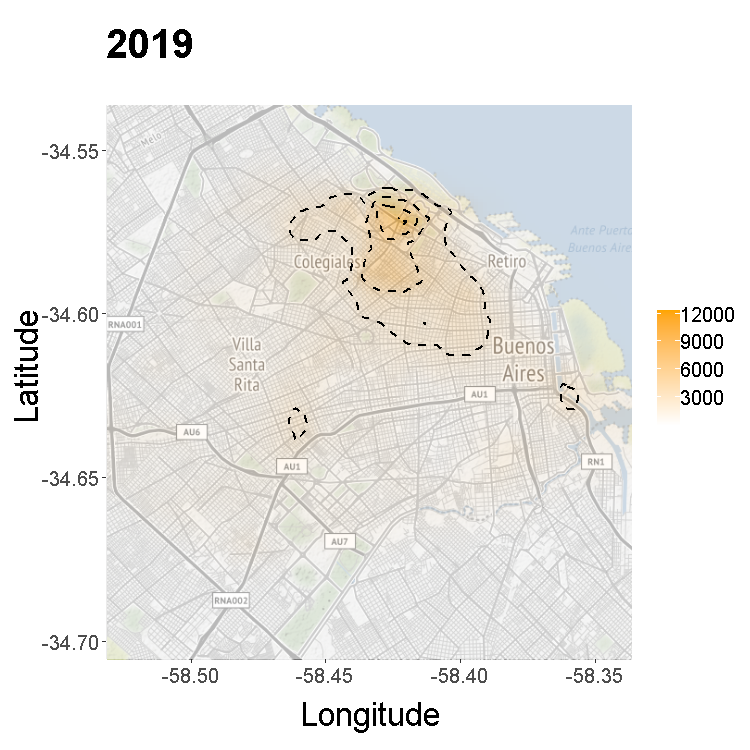}
    } 
        \subfloat[]
        {
        \includegraphics[scale=0.3]{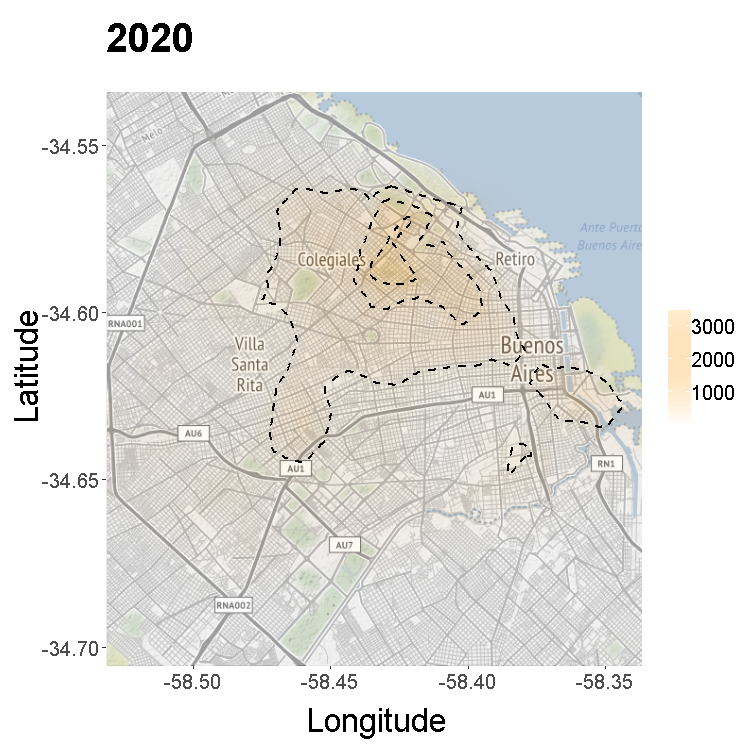}
        } \\
\end{tabular}
  \caption{(a) Fitted BMD corresponding to $c=8$km$^2$ along with a contour of pointwise differences between fitted posterior intensity functions over consecutive years. (b) Posterior density estimation corresponding to the centers of of the fitted BMD. (c) and (d) shows the fitted intensity functions for 2019 and 2020. \label{fig:TheftBA3}}
\end{figure}

In Fig.~\ref{fig:TheftBA3}(a) we depict the estimated BMD corresponding to an area of $8$km$^2$ along with an heat map of the differences in the estimated posterior intensity functions between consecutive years; the value of $8$km$^2$ was chosen for illustration as it corresponds to about twice the size of the largest neighborhood, which is Palermo. We also depict in Fig.~\ref{fig:TheftBA3}(b) an heat map of the posterior density corresponding to the center of the BMD which shows that these are substantially concentrated, thus suggesting low uncertainty on the fitted BMD.

To clarify the applied meaning of such BMDs we provide some additional background on the social context surrounding the empirical analysis. In Fig.~\ref{fig:TheftBA3}(c--d) we depict the fitted intensity functions corresponding to both years. As can be seen from Fig.~\ref{fig:TheftBA3}, in 2019 and 2020 thefts were more likely to occur in APRV (Almagro, Palermo, Recoleta, and Villa Crespo) which are some of the neighborhoods where several commercial and touristic activities took place. Yet, important differences on the estimated intensity functions are perceived between both years. During the first half of year 2020, local authorities took strong social distancing measures such as the limitation to the access the public transportation system, restrictions on business and commerce during the day, limitations on gatherings and tourism activities, restriction to the capacity in bars and restaurants, among others. The difference on the estimated intensity functions between consecutive years evident from Fig.~\ref{fig:TheftBA3}---and the implied reduction of thefts over 2020---is in line with the findings of  \cite{mohler2020impact} that report similar evidence on the effect of COVID-19 lockdown and social distance policies in nonviolent crime.

The BMD in Fig.~\ref{fig:TheftBA3}(a) suggests that APRV (Almagro, Palermo, Recoleta, and Villa Crespo) are the neighborhoods where there was a most impactful effect of COVID-19 lockdown. To put it differently, while nonviolent crime has decreased during lockdown over the entire city, what the fitted BMD highlights is that such reduction was \textit{relatively} much higher in the APRV neighborhoods. 



\subsection{Volatility in Stock Markets}\label{sec:vol}
Our second illustration will shed light on the multi-objective approach from Section~\ref{multi}. 
Data were gathered from \href{https://www.finance.yahoo.com}{Yahoo Finance} and consist of monthly values of the NYSE and NASDAQ composite indices from the New York Stock Exchange ($\{X_t\}$) and the NASDAQ Stock Exchange ($\{Y_t\}$), respectively. The data ranges from January 1990 to May 2021, thus covering a variety of episodes of financial turbulence such as the dotcom tech bubble that peaked around 2000, the subprime crisis that started around 2007, and the recent COVID--19 global pandemic. In the multi-objective RMD analysis to be conducted here, we will consider two functional parameters of interest: The mean values of the indices over time and the volatility of their log returns, that is, 
\begin{equation*}
\begin{cases}
  m_X(t) = E(X_t), \\
  m_Y(t) = E(Y_t), 
\end{cases}\qquad 
\begin{cases}
  \sigma_X(t) = [E\{\log(X_t/X_{t-1})^2\}]^{1/2}, \\
  \sigma_Y(t) = [E\{\log(Y_t/Y_{t-1})^2\}]^{1/2}.
\end{cases}
\end{equation*}
These functional parameters were modeled according to \eqref{gspec} exactly as in Section~\ref{simulation}, that is, using an identity link function and B-spline basis functions, choosing the number of basis functions using the DIC, and using the same uninformative prior.

We consider intervals of six months, one year, and two years (corresponding to $c = 6, 12$, and 24) and we aim to evaluate on what periods of such length these two stock markets differed the most---in terms of both average returns as well as volatility. We thus seek for the interval of time $B^{*}_c = [t_{c}^* - r_{c}^*, t_{c}^* + r_{c}^*]$ that as in \eqref{soll} maximizes the following scalarized set function optimization problem, 
\begin{equation}\label{eq14}
  \begin{array}{rl}
    {\max}\{\mathcal{F}_w\{B(t, r)\}: (t, r) \in T \times [0, R_c]\},
  \end{array}
\end{equation}
where $w$ is the scalarization parameter and
$$\mathcal{F}_w\{B(t, r)\} = w\int_{t - r}^{t + r} |m_X(u) - m_Y(u)| \, \dif u + (1-w) \int_{t - r}^{t + r}|\sigma_{X}(u) - \sigma_{Y}(u)|\, \dif u.$$
It follows from Theorem~\ref{scal} that every solution to the linear  scalarization problem \eqref{eq14} is a Pareto optimal BMMD (ball of multi-maximum dissimilarity), and hence Pareto optimal BMMDs obtained by linear scalarization have the nice feature of allowing one to put more emphasis on the mean values or on volatilities according to how we set $w$. That is, by setting $w=0$ or $w=1$, we only consider volatilities or mean values respectively and the analysis corresponds to a standard BMD, while for $w\in (0,1)$ absolute values in differences between mean functions are more important than those in volatilities as $w$ increases. In terms of computing we adapt Algorithm~\ref{alg} to the multi-objective context. That is, inference about the BMMDs is conducted by sampling $m = 1\,000$ times from the posteriors for means and volatilities---and rather than solving \eqref{maxb} as in Algorithm~\ref{alg}---we now solve the scalarized set function optimization problem in \eqref{eq14}.

\begin{figure}\vspace{-.8cm}
  \centering \footnotesize 
  \textbf{BMD for mean} $(w = 1)$\\
  \begin{tabular}{cc} 
    \subfloat[]{\includegraphics[scale=0.2]{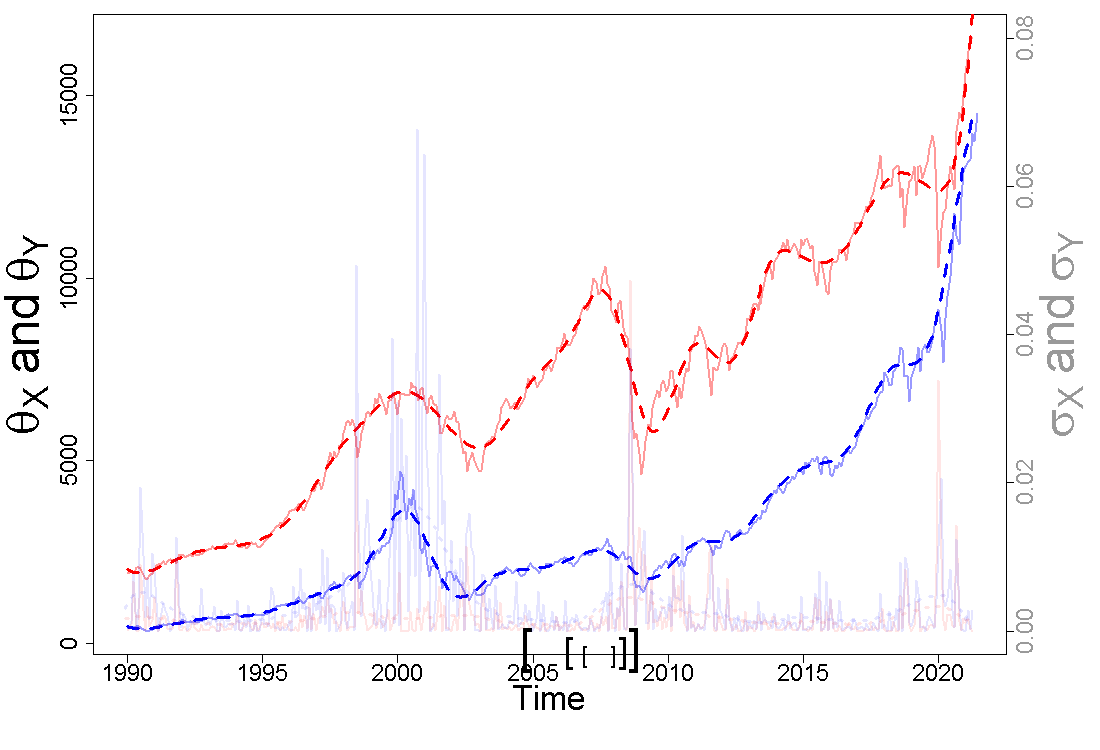}} & 
\subfloat[]{\includegraphics[scale=0.2]{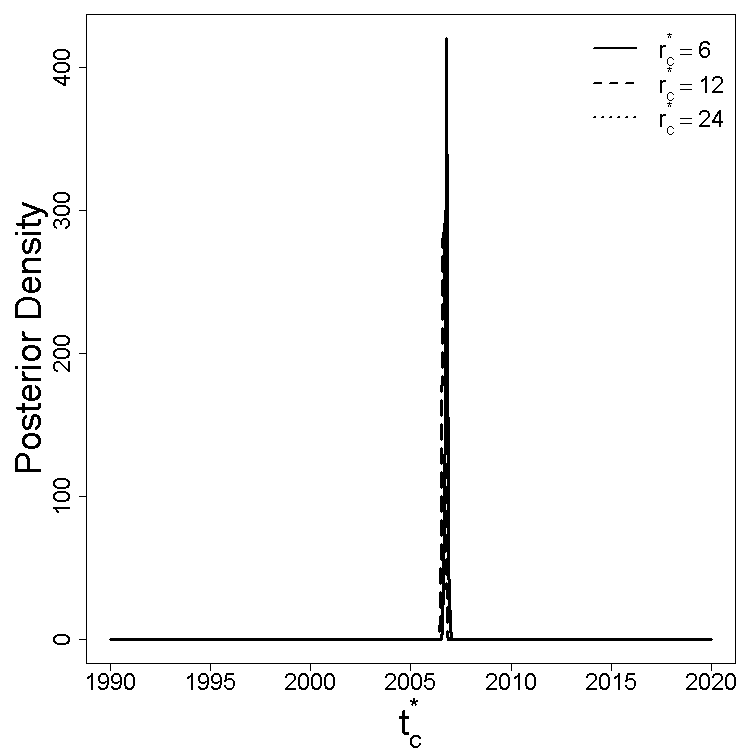}} \\
\end{tabular}~\\~\\
\textbf{BMD for volatility} $(w = 0)$
  \begin{tabular}{cc} 
\subfloat[]{\includegraphics[scale=0.2]{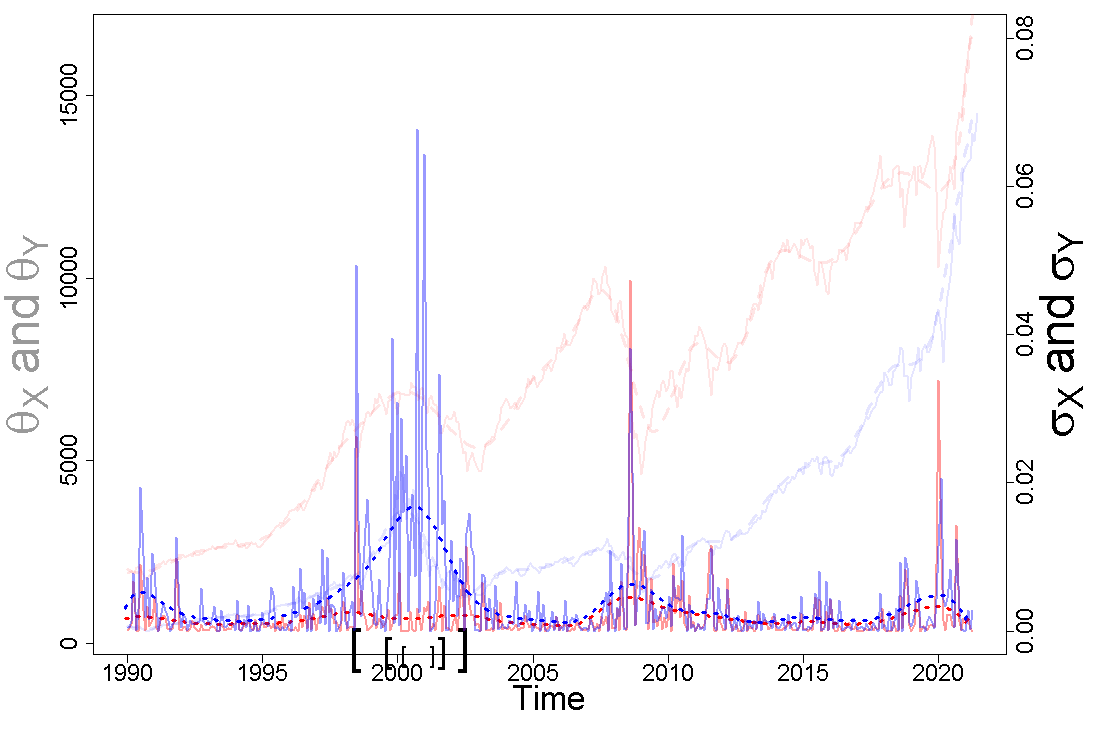}} &
\subfloat[]{\includegraphics[scale=0.2]{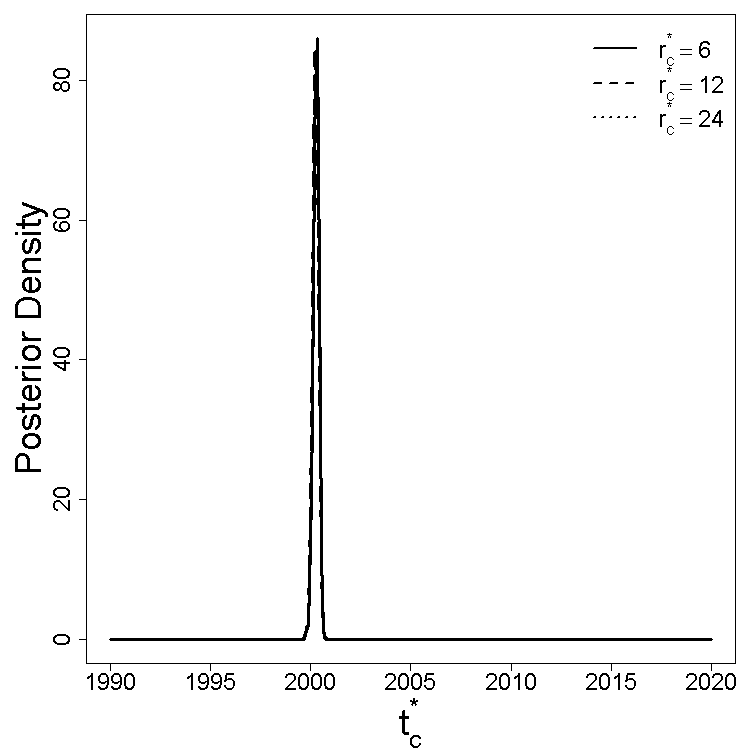}} \\
  \end{tabular}~\\~\\
  \textbf{Multi-objective BMD for mean--volatility} $(w = 2/3)$
    \begin{tabular}{cc} 
\subfloat[]{\includegraphics[scale=0.2]{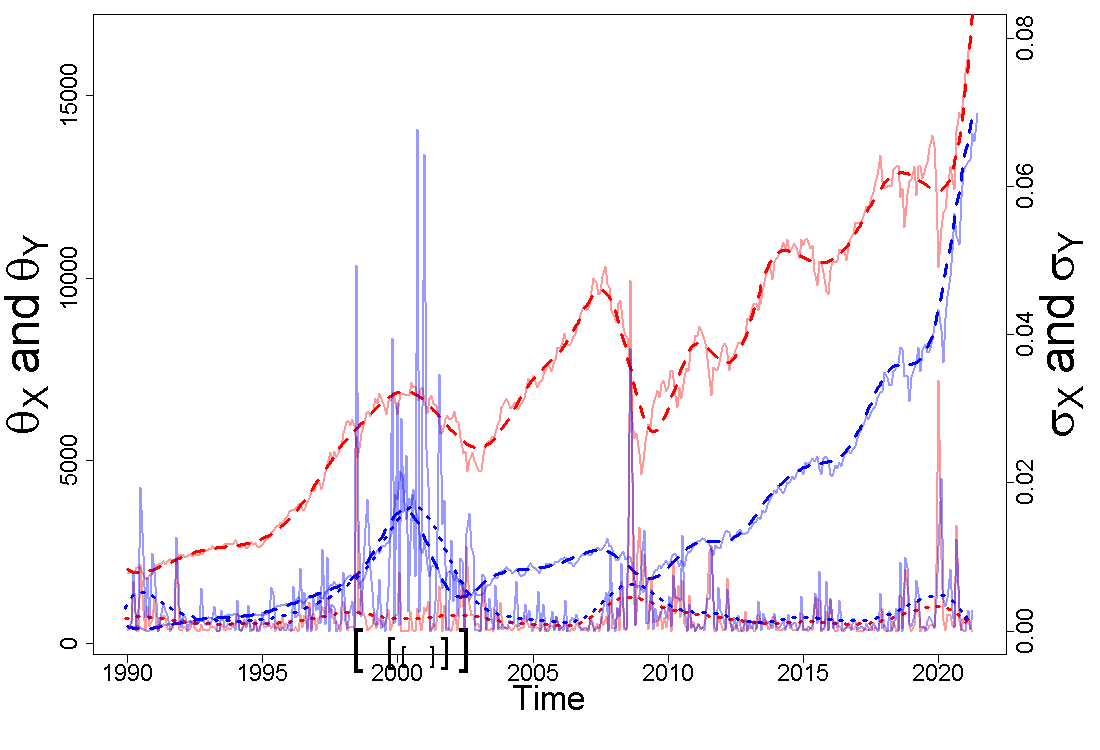}} &
\subfloat[]{\includegraphics[scale=0.2]{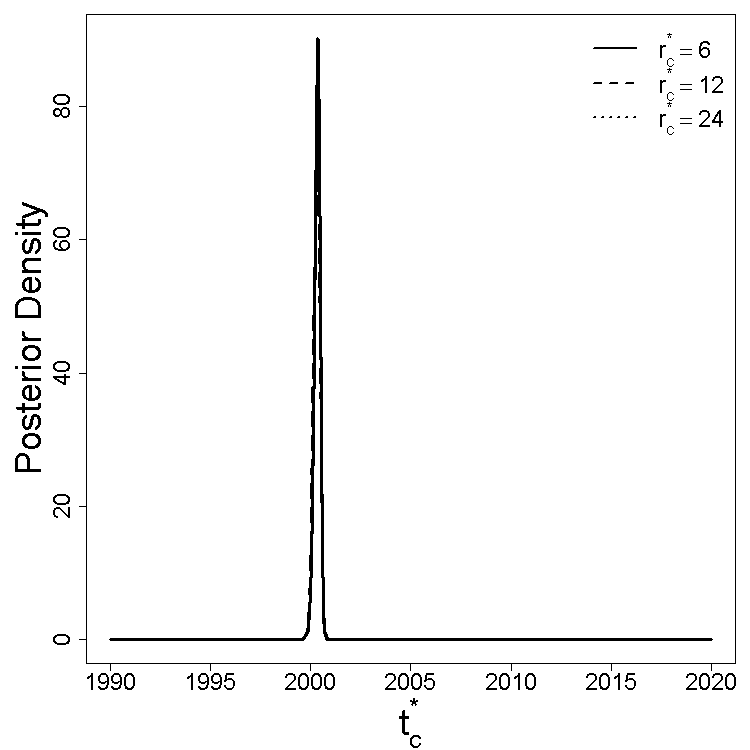}}
  \end{tabular}                                                                        
  \caption{(a, c, e) The brackets on the time axis correspond to the fitted multi-objective BMDs for six monts, one year, and two years for $w = 1$, $w = 0$, and $w = 2 / 3$; the charts also show the realized (solid) and estimated (dotted) volatilities corresponding to NYSE (red) and NASDAQ (blue). (b, d, f) Marginal posterior density for the centre of the BMD. \label{fig:volat}}
\end{figure}

In Fig.~\ref{fig:volat} we compare the results of the single BMD analysis [panels (a) to (d)] versus the multi-objective analysis [panels (e) and (f)] considering a value of $c$ corresponding to a period of six months, one year, and two years. As can be seen from Fig.~\ref{fig:volat}(a--b), the BMD associated with the mean ($w = 1$) concentrates around the subprime crisis, indicating that July 2006 to July 2008 is the period over which mean levels of NYSE and NASDAQ differed the most. In Fig.~\ref{fig:volat}(c--d), we see that the BMD associated with volatility ($w = 0$) ranges from October 1999 to October 2001---which corresponds to the dotcom bubble burst. Finally, the multi-objective approach with $w=2/3$ is depicted in Fig.~\ref{fig:volat} (e--f) and it suggests that volatility has a greater control on the objective function in \eqref{eq14}; that is, even when we set $w=2/3$---that is, even when we set more emphasis on the differences in means rather than the differences in volatility---we still get a similar result as setting $w = 0$, as we end up recovering the period of the dotcom bubble burst as can be seen from Fig.~\ref{fig:volat}(e--f).

\subsection{Electrocardiogram Data (ECG200)}\label{sec:ecg}
For our final illustration we use the ECG200 dataset contributed by \cite{olszewski2001}. The data is the result of monitoring electrical activity recorded during one heartbeat and it consists of 200 ECG signals sampled at 96 time instants, corresponding to 133 normal heartbeats ($\mathcal{D}_X$) and 67 myocardial infarction signals ($\mathcal{D}_Y$); the data are publicly available from the \href{http://www.cs.ucr.edu/~eamonn/time_series_data_2018/}{UCR Time Series Classification and Clustering} website.
 
In this illustration, the functional parameters of interest are the mean functions of ECG signals for both classes (normal heartbeat $\theta_X(t) = \E(X_t)$, and myocardial infarction $\theta_Y(t) = \E(Y_t)$) and one of the goals of the analysis is to track down periods, of a given length, over a cardiac cycle where the differences between the two classes is most pronounced. To model these functional parameters the Gaussian process prior specification in \eqref{gspec} was once more applied using an identity link, B-spline basis functions, the DIC to select the number of basis functions, and a Mat\'ern covariance function with the PC prior of \citet{fuglstad2019} setting $P(\sigma > 1) = 0.001$ and $P(\ell < 0.05) = 0.001$. 

\begin{figure}
  \centering
  \begin{tabular}{cc}
\subfloat[]{\includegraphics[scale=0.25]{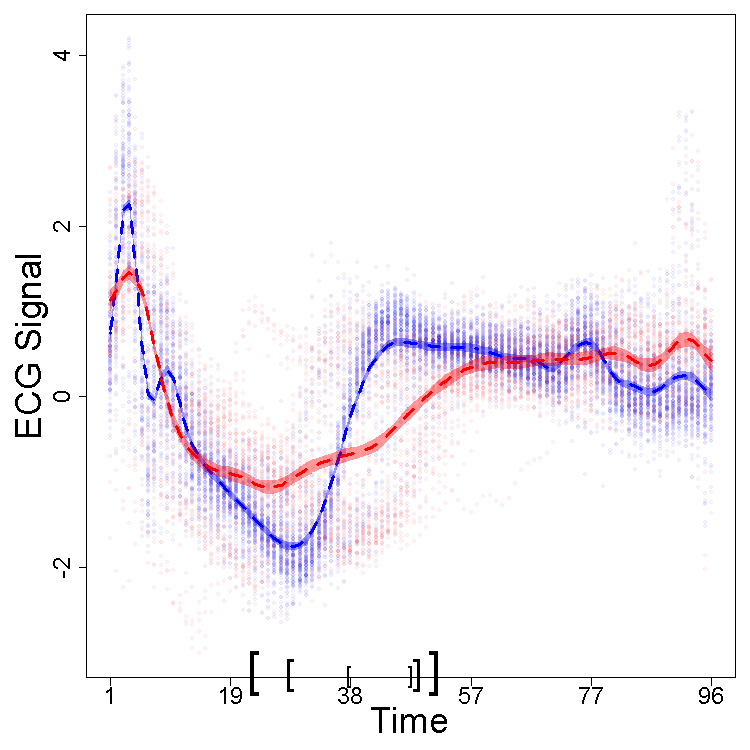}
} & \hspace{0.2cm}
\subfloat[]{\includegraphics[scale=0.25]{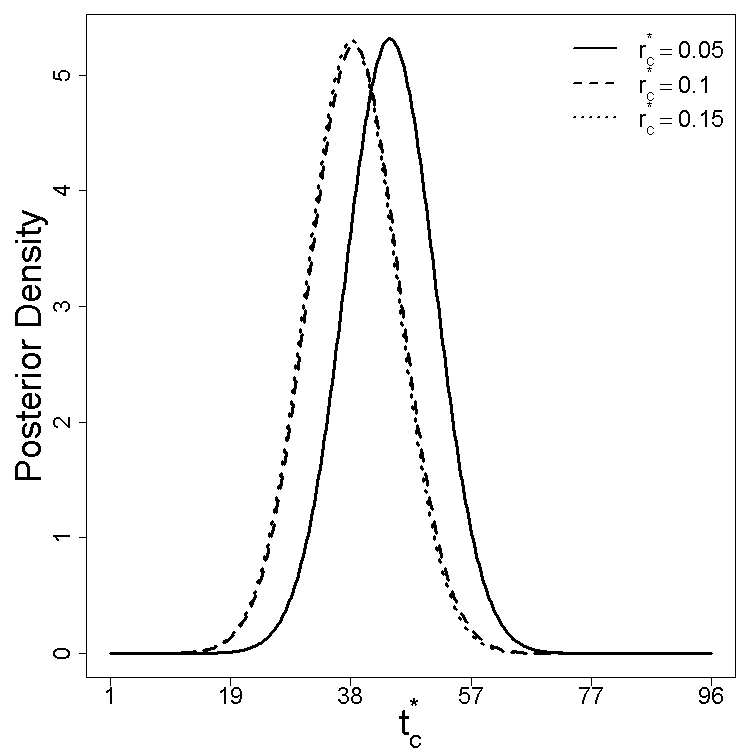}
}
\end{tabular}
\caption{ECG200 dataset: (a) The brackets on the time axis correspond to the fitted BMDs, $\widehat{B}^*_{10} \subset \widehat{B}^*_{20} \subset \widehat{B}^*_{30}$; the chart also shows the raw data (dots) and fitted mean functions (dashed) along with $95\%$ credible bands, where red and blue respectively correspond to normal heartbeats and  myocardial infarction signals. (b) Marginal posterior density for the centre of the set of maximum dissimilarity $B^*_{10}$. \label{fig:ecg1}}
\end{figure}

In Fig.~\ref{fig:ecg1}(a), we plot the fitted posterior estimates for a sequence of BMDs ($\widehat{B}^*_{10}$, $\widehat{B}^*_{20}$, and $\widehat{B}^*_{30}$) using brackets in the time axis, along with the raw data, and the fitted mean functions with 95\% credible bands. The obtained BMDs uncover periods of a given length where we observe largest differences between the estimated mean ECG functions. Informally, we may think of such BMDs as corresponding to intervals of about 10--30 deciseconds, since the 96 time instants cover about a cardiac cycle for all subjects and thus are not expected to last longer than 1 second. The fitted BMD centers are localized around the time instants 22--55. All in all, the analysis suggests that while normal heartbeats and myocardial infarction signals have similar `peaks' at the beginning of the sample period (i.e.~they have similar Q waves, in ECG signal analysis terminology), immediately in the period right after (i.e.~over their ST segments) they greatly differ. 

We close the analysis with two final remarks. First, in this illustration the fitted BMDs verify the chained inclusions $\widehat{B}^*_{10} \subset \widehat{B}^*_{20} \subset \widehat{B}^*_{03}$, but this property does not hold in general---nor for the fitted BMDs, nor for the true ones; counterexamples can be constructed either numerically or analytically. Second, although BMDs are unrelated to classification, since the ECG200 dataset is a popular benchmark for new classifiers it may be sensible to ask whether the accuracy of some classifiers at discriminating outcome classes (diseased--nondiseased) can be improved by focusing on BMDs rather than treating the entire time horizon equally; we leave such open problem for future analysis. 

\section{Closing Remarks}\label{discussion}
Regions of maximum dissimilarity and their variants are here proposed as
a tool for acquiring knowledge on the region with a given size where two stochastic processes differ the most.
The proposed learning problem is shown to be equivalent to a \textit{continuous set
function optimization} on a \textit{monotone modular} function, under a \textit{Lebesgue
measure constraint}. As a byproduct, the paper contributes to the
literature on set function optimization which thus far is focused on a
discrete and combinatorial framework, and which has been focused on
monotone submodular functions \cite[e.g.][]{nemhauser1978,
  calinescu2011, goldengorin2009, buchbinder2017, buchbinder2018},
typically under cardinality or matroid constraints.

The existence of the proposed regions of maximum dissimilarity is
nontrivial but we prove their existence, and illustrate with artificial
and real data that it only requires a moderate computational
investment to learn them from data. The proposed methods are
developed in full generality for the setting where the data of
interest are themselves stochastic processes, and thus the proposed
toolbox can be used for unveiling the regions of maximum dissimilarity
with a given volume, for a variety of random process data.
All modeling was framed within a latent Gaussian
framework, with inference being conducted using the 
Integrated Nested Laplace Approximation; clearly, other computational
approaches could have been employed as well, including, for example, variational inference
\citep{blei2017}.

A multi-objective version of the proposed framework is also devised 
so to learn about multi-objective RMDs, where several 
functional parameters are considered---each characterizing a specific 
feature of the processes being compared.
In addition, another variant of the method to which we refer to as
Hardy--Littlewood BMDs showcases that the current framework includes
the Youden index and the Youden's $\text{J}$ statistic as well as the
Kolmogorov metric as particular cases. 

While the theoretical developments from Section~\ref{sec:dis} establish
the existence of general compact and convex sets of maximum dissimilarity,
BMDs turn out be a much convenient simplification for a variety of reasons.
First, numerical optimization is much more challenging with general RMDs, 
whereas for BMDs it is relatively simple as can be seen from \eqref{maxb}.   
Second, the inference for general RMDs would entail averaging 
posterior simulated RMDs---that is, averaging sets---whereas with BMDs we just need to
compute posterior mean of the $(d+1)$-dimensional centre-radius pair.

\textbf{Acknowledgments}. We thank, without implicating, Vanda In\'acio de Carvalho for comments
and feedback and Finn Lindgren for discussions on INLA. MdC was
partially supported by FCT (Funda\c c\~ao para a Ci\^encia e a
Tecnologia, Portugal) through the projects PTDC/MAT-STA/28649/2017 and
UID/MAT/00006/2019.

\appendix
\section{Technical Details and Auxiliary Lemmata}\label{lemmata}
In this section we state some auxiliary facts that will be used to
prove the main results of this paper.
Beyond the auxiliary lemmata to be stated below we use some basic facts 
from measure, topology, and convex analysis, such as, for example,
Lebesgue differentiation theorem
\citep[e.g.][Theorem~1.6.19]{tao2011}, Tikhonov's theorem
\citep[][Theorem~5.3.1]{waldmann2014}, and the well-known fact that
the volume functional is continuous in the space of convex bodies 
\citep[][Theorem~1.8.16]{schneider2014}, under the Hausdorff metric.

Recall
that Tikhonov's theorem implies that from the Cartesian product of two compact sets
results a compact set. In addition, recall that
Lebesgue differentiation theorem implies that if $f:\mathbf{R} \to \mathbf{C}$ is an absolutely integrable function,
then for almost every $x \in \mathbf{R}^d$,
\begin{equation*}
  \lim_{r \to 0} \frac{1}{|B(x, r)|} \int_{B(x, r)} f(y) \, \dif y = f(x), 
\end{equation*}
where $B(x, r) = \{y: \|x -  y\|_1 \leq r\}$. 

We now present the auxiliary lemmata. Blaschke selection theorem is a
classical result on convex analysis; the version stated below can be
found, for instance, in \cite{benyamini1998}. In addition, we also recall below
two key results on optimization of correspondences (i.e.~set-valued
functions), namely Berge's maximum theorem and the product of correspondences theorem; 
the versions below can be found in 
\citet[][Theorems~17.31 and 17.28]{aliprantis2006}.

\begin{lemma}[Blaschke selection theorem]\label{blaschket}
 The set of all compact convex subsets of a fixed compact subset of
$\mathbf{R}^d$ is compact under the Hausdorff metric.
\end{lemma}

\begin{lemma}[Berge's maximum theorem]\label{berget}
  Let $\varphi: X \twoheadrightarrow Y$ be a continuous correspondence
  between topological spaces with nonempty compact values, and suppose
  that \linebreak $g: \text{Gr } \varphi \to \mathbb{R}$ is continuous. Define
  the ``value function'' $v: X \to \mathbf{R}$ by
  \begin{equation*}
    v(x) = \max\{g(x, y): y \in \varphi(x)\}  
  \end{equation*}
  and the correspondence $\alpha: X \twoheadrightarrow Y$ of maximizers by
  \begin{equation*}
    \alpha(x) = \{y \in \varphi(x): g(x, y) = v(x)\}.
  \end{equation*}
  Then:
  \begin{enumerate}
  \item The value function $v$ is continuous.
  \item The ``argmax'' correspondence $\alpha$ has nonempty compact values.
  \item If $Y$ is Hausdorff, then the ``argmax'' correspondence $\alpha$ is upper hemicontinuous.  
  \end{enumerate}
\end{lemma}

\begin{lemma}[Product of correspondences theorem]\label{prodt}
  The product of correspondences obeys the following properties:
  \begin{enumerate}
  \item The product of a family of upper hemicontinuous correspondences with compact values
    is upper hemicontinuous with compact values.
  \item The product of a finite family of lower hemicontinuous
    correspondences is lower hemicontinuous.
  \end{enumerate}
\end{lemma}

\section{Proofs of Main Results}\label{proofs}
\subsection{Proof of Theorem~\ref{props}}\label{Proof}


\textbf{Claim~1}.~We start by showing that $$\mathscr{F}_c = \{A \in \mathcal{A}: |A| \leq c\}$$ is compact, for every $c \geq 0$ where $\mathcal{A}$ is the family of compact and convex subsets of the ground set $T$. Recall that by the Blaschke selection theorem (Lemma~\ref{blaschket}), $\mathcal{A}$ is compact under the Hausdorff metric. Further, since the volume functional $|\cdot|$ is continuous \citep[][Theorem~1.8.16]{schneider2014}, and given that $\mathscr{F}_c$ is the preimage of the closed set $[0, c]$, it follows that $\mathscr{F}_c$ is a closed subset of $\mathcal{A}$ and hence it is compact.
    
Observe next that maximizing $F(A) = \|\theta_X - \theta_Y\|^{(A)}_p$ is equivalent to maximizing $F^{p}(A)$, for $p > 0$, and as we show next $F^{p}(A)$ is upper semicontinuous under the Hausdorff metric, for every $A \in \mathscr{F}_c$. Let $A_n \to A$ in $(\mathscr{F}_c, d_H)$, for a fixed $c \geq 0$. It can be easily shown that \cite[e.g.][Theorem 12.3.6]{schneider2008} 
    \begin{equation}\label{1at}
      1_{A}(t) \geq \lim \underset{n}{\sup} \, 1_{A_n}(t), \quad
      t \in T \subset \mathbf{R}^d.
    \end{equation}
    Combining \eqref{1at} with Fatou's lemma yields that 
  \begin{equation*}
    \begin{split}
      F^{p}(A) &= \int_T 1_A(t) |\theta_X(t) - \theta_Y(t)|^p \, \mu(\dif t) \\
      &\geq \int_T \lim \underset{n}{\sup}\, 1_{A_n}(t) |\theta_X(t) - \theta_Y(t)|^p \, \mu(\dif t) \\
      &\geq \lim \underset{n}{\sup} \int_T 1_{A_n}(t) |\theta_X(t) - \theta_Y(t)|^p \, \mu(\dif t) \\
      &= \lim \underset{n}{\sup}\, F^{p}(A_n),
    \end{split}
    \end{equation*}
    thus showing that $F^{p}(A)$ is upper semicontinuous under the Hausdorff metric, for every $A \in \mathscr{F}_c$. The final step of the proof is tantamount to a standard argument used for proving Weierstrass theorem. Let $u_c = \sup\{F^{p}(A): A \in \mathscr{F}_c\} \cup \{\infty\}$. By definition, for every $c$ there exists a maximizing sequence $A_{n} \in \mathscr{F}_c$ such that $F^{p}(A_{n}) \to u_c$. By compactness, we can assume that $A_{n} \to A^*_c$. Upper semicontinuity of $F^{p}(A)$ implies that
    $u_c = \lim {\sup}_n\, F^{p}(A_{n}) \leq F^{p}(A^*_c),$ and on the other hand we have $u_c \geq F(A^*_c)$ since $u_c$ is the supremum. This proves that $u_c = F^{p}(A^*_c)$ is the maximum of $F^{p}(A)$, subject to $A \in \mathscr{F}_c$, and hence $A^*_c$ exists that solves \eqref{eq:problem}.\\

\noindent \textbf{Claim~2}.~First, observe that 
  \begin{equation*}
    \begin{split}
      \|K\theta_X - K\theta_Y\|_p^{(A)} &= \| \alpha + \beta \theta_X - (\alpha + \beta \theta_Y) \|_p^{(A)} \\
      &= |\beta| \times \|\theta_X-\theta_Y\|_p^{(A)},
    \end{split}
  \end{equation*}
  from where it follows that a set $A$ that maximizes $\|K\theta_X - K\theta_Y\|_p^{(A)}$ also maximizes \linebreak  $\|\theta_X-\theta_Y\|_p^{(A)}$; that is, $A^K_c = A^*_c$ for all $c \in [0, \infty)$. Second, it follows from the change of variables formula that 
  \begin{equation*}
    \begin{split}
      \|L \theta_X - L \theta_Y\|_p^{(A^L)} &=
      \bigg[ \int_{A^{L}} \{\theta_X(\alpha + \beta t) - \theta_Y(\alpha + \beta t)\}^{p} \mu(\dif t) \bigg]^{1/p}\\
      &= \frac{1}{\beta^{1/p}} \bigg[\int_A \{\theta_X(u) - \theta_X(u)\}^{p} \mu(\dif u)\bigg]^{1/p},
    \end{split}
  \end{equation*}
  where $A = \{\alpha + \beta t: t \in A^{L}\}$. Thus,
  \begin{equation*}
    \arg \max\{\|L \theta_X - L \theta_Y\|_p^{(A^L)}: A^L \in \mathscr{F}_c\} =
    \arg \max\{\|\theta_X - \theta_Y\|_p^{(A)}: A \in \mathscr{F}_{\alpha + \beta c}\},
  \end{equation*}
  and hence, $A^L_c = A^*_{\alpha + \beta c}$. \\
  
\noindent \textbf{Claim~3}.~Note first that $D_c^{*} \geq 0$, for all $0 < c < \infty$. Also, it holds that $D_c^{*} = 0$ if and only if $\theta_X = \theta_Y$ in $\Theta_c$; indeed, $D_c^{*} = 0$ implies that $0 = \|\theta_X - \theta_Y\|^{(A^*_c)}_p \geq  \|\theta_X - \theta_Y\|^{(A)}_p$,  for every $A \in \mathscr{F}^c$, which is only possible if $\theta_X = \theta_Y$ in $\Theta_c$, as $c > 0$. Finally, the triangle inequality for $\|\cdot \|_p^{(A)}$ and $\|F\|_\infty = \max_A  |F(A)|$ yields that 
    \begin{equation*}
      \begin{split}
        D^*_c(\theta_X, \theta_Z) &= \max\{\|\theta_X - \theta_Z\|^{(A)}_p: A \in \mathscr{F}_c\} \\ 
        &= \max\{\|(\theta_X - \theta_Y) + (\theta_Y - \theta_Z)\|^{(A)}_p: A \in \mathscr{F}_c\} \\
        &\leq \max\{\|\theta_X - \theta_Y\|^{(A)}_p + \|\theta_Y - \theta_Z)\|^{(A)}_p: A \in \mathscr{F}_c\} \\
        &\leq \max\{\|\theta_X - \theta_Y\|^{(A)}_p : A \in \mathscr{F}_c\} +
        \max\{\|\theta_Y - \theta_Z)\|^{(A)}_p: A \in \mathscr{F}_c\} \\
        &=  D^*_c(\theta_X, \theta_Y) + D^*_c(\theta_Y, \theta_Z),
      \end{split}
    \end{equation*}
    hence concluding the proof.\\

\noindent \textbf{Claim~4}.~First, we note that an increase in $c$ represents augmenting the search domain as $$\mathscr{F}_a \subset \mathscr{F}_b, \quad \text{for } b > a.$$ This combined with the fact that the set objective function $F(A) =  \|\theta_X - \theta_Y\|_p^{(A)}$ is non-decreasing ($A \subseteq B$ implies $F(A) \leq F(B)$) yields that $D^*_b \geq D^*_a$, for $b > a$.

  
\subsection{Proof of Theorem~\ref{Dc2}}
\textbf{Claim~1.~}As a consequence of Tikhonov's theorem (Appendix~\ref{lemmata}), the search domain $T \times [0, R_c]$ is compact for every $c \geq 0$. Next, it is a routine exercise to prove that $f(t, r) = F\{B(t, r)\}$, is continuous for all $(t, r) \in T \times [0, R_c]$ as both $T$ and $[0, R_c]$ are compact. The final result then follows from Weierstrass theorem.\\

\noindent \textbf{Claim~2.~}Let $\theta_X, \theta_Y \in L^p(T)$ be fixed and set $\mathscr{D}^{*}_c = \mathscr{D}^{*}_c(\theta_X, \theta_Y).$ The proof uses Berge's maximum theorem (Lemma~\ref{berget}) which, as can be seen from Appendix~\ref{lemmata}, claims that a value function is continuous provided that both the objective function and the constraint correspondence are continuous. In our setup, the value function is
\begin{equation*}
  \begin{split}
  \mathscr{D}^{*}_c &= \max \{g(c, t, r): (t, r) \in \varphi(c)\}\\
  &= \max \{f(t, r): (t, r) \in \varphi(c)\},\\    
  \end{split}
\end{equation*}
where $g(c, t, r) = f(t, r) = F\{B(t, r)\}$, and the constraint correspondence is $\varphi:[0, \infty) \twoheadrightarrow \mathbf{R}^{d + 1}$, defined by
    \begin{equation}
      \label{varphi}
      \varphi(c) = T \times [0, R_c], \qquad c \geq 0.
    \end{equation}
    Thus, we just have to prove that $\varphi(c)$ in \eqref{varphi} is continuous, given that the objective function $g(c, t, r) = f(t, r) = F\{B(t, r)\}$ is trivially continuous for all $(t, r) \in T \times [0, R_c]$. Continuity of $\varphi(c)$ follows immediately from the product of correspondences theorem (Lemma~\ref{prodt}), which yields that $\varphi(c) = \varphi_1(c) \times \varphi_2(c)$ is continuous as both $\varphi_1(c) = T$ and $\varphi_2(c) = [0, R_c]$ are compacted-valued, and $R_c$ is a continuous function for every $c \geq 0$. Finally, upper semicontinuity of the argmax correspondence,
    \begin{equation*}
      \begin{split}
        \alpha_c &= \{(t, r) \in T \times [0, R_c]: g(c, t, r) = \mathscr{D}_c^{*}\} \\
        &= \{(t, r) \in T \times [0, R_c]: f(t, r) = \mathscr{D}_c^{*}\}, \\
      \end{split}      
    \end{equation*}
    follows also from Berge's maximum theorem as $\mathbf{R}^{d + 1}$ is Hausdorff.

\subsection{Proof of Theorem~\ref{scal}}
Suppose by contradiction that $A'_{w, c}$ is the solution to the set function linear scalarization problem \eqref{soll}, for a fixed $w \in (0, \infty)^q$, but that $A'_{w, c}$ was not a Pareto optimal RMMD; then, there would exist a Pareto improvement $A \in \mathscr{F}_c$ , with $A \neq A'_{w, c}$, so that $F_i(A) \geq F_i(A'_{w, c})$, for all $i$, and $F_i(A) > F_i(A'_{w, c})$, for at least an $i$. But then, $$\sum_{i = 1}^q w_i F_i(A) > \sum_{i = 1}^q w_i F_i(A'_{w, c}),$$ which is a contradiction as $A'_{w, c}$ solves the set function linear scalarization problem.

\subsection{Proof of Proposition~\ref{dhlprop}}
  The proof of the first claim is straightforward. For every $c > 0$, 
    \begin{equation*}
      \begin{split}
      \mathbb{D}_c^{*}(\theta_X, \theta_Y) &=
      \max\bigg\{\dashint_{B(t, r)} |\theta_X(u) - \theta_Y(u)|  \, \dif u: (t, r) \in T \times [0, R_c]\bigg\} \\
      &= \max \bigg\{\frac{1}{|B(t, r)|}\int_{B(t, r)} |\theta_X(u) - \theta_Y(u)|  \, \dif u: (t, r) \in T \times [0, R_c]\bigg\} \\
      &\leq  \max \bigg\{\frac{1}{\cancel{|B(t, r)|}}\cancel{|B(t, r)|} \max_{u \in T}|\theta_X(u) - \theta_Y(u)|: (t, r) \in T \times [0, R_c]\bigg\} \\
      &= \max_{t \in T}|\theta_X(t) - \theta_Y(t)|.
    \end{split}
  \end{equation*}
  The second claim follows directly from the Lebesgue differentiation theorem (Appendix~\ref{lemmata}).









\vskip 0.2in
\bibliography{library}

\begin{thebibliography}{50}
\providecommand{\natexlab}[1]{#1}
\providecommand{\url}[1]{\texttt{#1}}
\expandafter\ifx\csname urlstyle\endcsname\relax
  \providecommand{\doi}[1]{doi: #1}\else
  \providecommand{\doi}{doi: \begingroup \urlstyle{rm}\Url}\fi

\bibitem[Abramowitz and Stegun(1964)]{abramowitz1964}
Milton Abramowitz and Irene~A Stegun.
\newblock \emph{Handbook of Mathematical Functions}.
\newblock Dover, New York, 1964.

\bibitem[Aliprantis and Border(2006)]{aliprantis2006}
CD~Aliprantis and KC~Border.
\newblock \emph{Infinite Dimensional Analysis}.
\newblock Springer, New York, 2006.

\bibitem[Benyamini(1998)]{benyamini1998}
Yoav Benyamini.
\newblock Applications of the universal surjectivity of the cantor set.
\newblock \emph{The American Mathematical Monthly}, 105\penalty0 (9):\penalty0
  832--839, 1998.

\bibitem[Berrendero et~al.(2016)Berrendero, Cuevas, and
  Torrecilla]{berrendero2016}
Jos{\'e}~R Berrendero, Antonio Cuevas, and Jos{\'e}~L Torrecilla.
\newblock Variable selection in functional data classification: A
  maxima-hunting proposal.
\newblock \emph{Statistica Sinica}, pages 619--638, 2016.

\bibitem[Berrendero et~al.(2020)Berrendero, Bueno-Larraz, and
  Cuevas]{berrendero2020}
Jos{\'e}~R Berrendero, Beatriz Bueno-Larraz, and Antonio Cuevas.
\newblock On {M}ahalanobis distance in functional settings.
\newblock \emph{Journal of Machine Learning Research}, 21\penalty0
  (9):\penalty0 1--33, 2020.

\bibitem[Blangiardo and Cameletti(2015)]{blangiardo2015}
Marta Blangiardo and Michela Cameletti.
\newblock \emph{Spatial and Spatio-temporal Bayesian Models with R-INLA}.
\newblock Wiley, New York, 2015.

\bibitem[Blei et~al.(2017)Blei, Kucukelbir, and McAuliffe]{blei2017}
David~M Blei, Alp Kucukelbir, and Jon~D McAuliffe.
\newblock Variational inference: A review for statisticians.
\newblock \emph{Journal of the American statistical Association}, 112\penalty0
  (518):\penalty0 859--877, 2017.

\bibitem[Borchers(2021)]{pracma}
Hans~W. Borchers.
\newblock \emph{pracma: Practical Numerical Math Functions}, 2021.
\newblock URL \url{https://cran.r-project.org/web/packages/pracma/pracma.pdf}.
\newblock R package version 2.3.6.

\bibitem[Buchbinder and Feldman(2018)]{buchbinder2018}
Niv Buchbinder and Moran Feldman.
\newblock Submodular functions maximization problems.
\newblock In \emph{Handbook of Approximation Algorithms and Metaheuristics, 2nd
  ed}, pages 753--788. Chapman and Hall/CRC, Boca Raton, FL, 2018.

\bibitem[Buchbinder et~al.(2017)Buchbinder, Feldman, and
  Schwartz]{buchbinder2017}
Niv Buchbinder, Moran Feldman, and Roy Schwartz.
\newblock Comparing apples and oranges: Query trade-off in submodular
  maximization.
\newblock \emph{Mathematics of Operations Research}, 42\penalty0 (2):\penalty0
  308--329, 2017.

\bibitem[Calinescu et~al.(2011)Calinescu, Chekuri, Pal, and
  Vondr{\'a}k]{calinescu2011}
Gruia Calinescu, Chandra Chekuri, Martin Pal, and Jan Vondr{\'a}k.
\newblock Maximizing a monotone submodular function subject to a matroid
  constraint.
\newblock \emph{SIAM Journal on Computing}, 40\penalty0 (6):\penalty0
  1740--1766, 2011.

\bibitem[Dette and Kokot(2021)]{dette2020}
Holger Dette and Kevin Kokot.
\newblock {Bio-equivalence tests in functional data by maximum deviation}.
\newblock \emph{Biometrika}, 108\penalty0 (4):\penalty0 895--913, 2021.
\newblock ISSN 0006-3444.
\newblock \doi{10.1093/biomet/asaa096}.
\newblock URL \url{https://doi.org/10.1093/biomet/asaa096}.

\bibitem[Faouzi and Janati(2020)]{faouzi2020}
Johann Faouzi and Hicham Janati.
\newblock pyts: A python package for time series classification.
\newblock \emph{Journal of Machine Learning Research}, 21\penalty0
  (46):\penalty0 1--6, 2020.
\newblock URL \url{http://jmlr.org/papers/v21/19-763.html}.

\bibitem[Ferraty and Vieu(2006)]{ferraty2006}
Fr{\'e}d{\'e}ric Ferraty and Philippe Vieu.
\newblock \emph{Nonparametric Functional Data Analysis: Theory and Practice}.
\newblock {Springer}, New York, 2006.

\bibitem[Fuglstad et~al.(2019)Fuglstad, Simpson, Lindgren, and
  Rue]{fuglstad2019}
Geir-Arne Fuglstad, Daniel Simpson, Finn Lindgren, and H{\aa}vard Rue.
\newblock Constructing priors that penalize the complexity of gaussian random
  fields.
\newblock \emph{Journal of the American Statistical Association}, 114\penalty0
  (525):\penalty0 445--452, 2019.

\bibitem[Goldengorin(2009)]{goldengorin2009}
Boris Goldengorin.
\newblock Maximization of submodular functions: Theory and enumeration
  algorithms.
\newblock \emph{European Journal of Operational Research}, 198\penalty0
  (1):\penalty0 102--112, 2009.

\bibitem[G{\'o}mez-Rubio(2020)]{gomez2020}
Virgilio G{\'o}mez-Rubio.
\newblock \emph{Bayesian Inference with INLA}.
\newblock Chapman and Hall/CRC, address={Boca Raton, FL}, 2020.

\bibitem[Gretton et~al.(2012)Gretton, Borgwardt, Rasch, Sch{\"o}lkopf, and
  Smola]{gretton2012kernel}
Arthur Gretton, Karsten~M Borgwardt, Malte~J Rasch, Bernhard Sch{\"o}lkopf, and
  Alexander Smola.
\newblock A kernel two-sample test.
\newblock \emph{Journal of Machine Learning Research}, 13\penalty0
  (1):\penalty0 723--773, 2012.

\bibitem[Horv{\'a}th and Kokoszka(2012)]{horvath2012}
Lajos Horv{\'a}th and Piotr Kokoszka.
\newblock \emph{Inference for {{Functional Data}} with {{Applications}}}.
\newblock {Springer}, New York, 2012.

\bibitem[{In{\'a}cio de Carvalho} et~al.(2017){In{\'a}cio de Carvalho}, {de
  Carvalho}, and Branscum]{inaciodecarvalho2017}
V.~{In{\'a}cio de Carvalho}, M.~{de Carvalho}, and A.~J. Branscum.
\newblock Nonparametric {{Bayesian}} covariate-adjusted estimation of the
  {{Youden}} index.
\newblock \emph{Biometrics}, 73\penalty0 (4):\penalty0 1279--1288, 2017.

\bibitem[Krainski et~al.(2018)Krainski, G{\'o}mez-Rubio, Bakka, Lenzi,
  Castro-Camilo, Simpson, Lindgren, and Rue]{krainski2018}
Elias Krainski, Virgilio G{\'o}mez-Rubio, Haakon Bakka, Amanda Lenzi, Daniela
  Castro-Camilo, Daniel Simpson, Finn Lindgren, and H{\aa}vard Rue.
\newblock \emph{Advanced Spatial Modeling with Stochastic Partial Differential
  Equations using R and INLA}.
\newblock Chapman and Hall/CRC, Boca Raton, FL, 2018.

\bibitem[Krause(2010)]{krause2010}
Andreas Krause.
\newblock {SFO}: A toolbox for submodular function optimization.
\newblock \emph{Journal of Machine Learning Research}, 11\penalty0
  (38):\penalty0 1141--1144, 2010.
\newblock URL \url{http://jmlr.org/papers/v11/krause10a.html}.

\bibitem[Lindgren et~al.(2015)Lindgren, Rue, et~al.]{lindgren2015}
Finn Lindgren, H{\aa}vard Rue, et~al.
\newblock Bayesian spatial modelling with {R-INLA}.
\newblock \emph{Journal of Statistical Software}, 63\penalty0 (19):\penalty0
  1--25, 2015.

\bibitem[Martins et~al.(2013)Martins, Simpson, Lindgren, and Rue]{martins2013}
Thiago~G Martins, Daniel Simpson, Finn Lindgren, and H{\aa}vard Rue.
\newblock Bayesian computing with {INLA}: New features.
\newblock \emph{Computational Statistics \& Data Analysis}, 67:\penalty0
  68--83, 2013.

\bibitem[Martos and {de Carvalho}(2018)]{martos2018}
G.~Martos and M.~{de Carvalho}.
\newblock Discrimination surfaces with application to region-specific brain
  asymmetry analysis.
\newblock \emph{Statistics in Medicine}, 11\penalty0 (37):\penalty0 1859--1873,
  2018.

\bibitem[Mohler et~al.(2020)Mohler, Bertozzi, Carter, Short, Sledge, Tita,
  Uchida, and Brantingham]{mohler2020impact}
George Mohler, Andrea~L Bertozzi, Jeremy Carter, Martin~B Short, Daniel Sledge,
  George~E Tita, Craig~D Uchida, and P~Jeffrey Brantingham.
\newblock Impact of social distancing during {COVID-19} pandemic on crime in
  {Los Angeles} and {Indianapolis}.
\newblock \emph{Journal of Criminal Justice}, 68:\penalty0 101692, 2020.

\bibitem[Nemhauser et~al.(1978)Nemhauser, Wolsey, and Fisher]{nemhauser1978}
George~L Nemhauser, Laurence~A Wolsey, and Marshall~L Fisher.
\newblock An analysis of approximations for maximizing submodular set
  functions—{I}.
\newblock \emph{Mathematical Programming}, 14\penalty0 (1):\penalty0 265--294,
  1978.

\bibitem[Nocedal and Wright(2006)]{nocedal2006}
Jorge Nocedal and Stephen Wright.
\newblock \emph{Numerical Optimization}.
\newblock Springer, New York, 2006.

\bibitem[Olszewski(2001)]{olszewski2001}
Robert~T Olszewski.
\newblock Generalized feature extraction for structural pattern recognition in
  time-series data.
\newblock Technical report, Carnegie-Mellon University, School of Computer
  Science, 2001.

\bibitem[Pardalos et~al.(2017)Pardalos, {\v{Z}}ilinskas, and
  {\v{Z}}ilinskas]{pardalos2017non}
Panos~M Pardalos, Antanas {\v{Z}}ilinskas, and Julius {\v{Z}}ilinskas.
\newblock \emph{Non-convex Multi-objective Optimization}.
\newblock Springer, New York, 2017.

\bibitem[Pini and Vantini(2016)]{pini2016}
Alessia Pini and Simone Vantini.
\newblock The interval testing procedure: {A} general framework for inference
  in functional data analysis.
\newblock \emph{Biometrics}, 72\penalty0 (3):\penalty0 835--845, 2016.

\bibitem[Pini and Vantini(2017)]{pini2017}
Alessia Pini and Simone Vantini.
\newblock Interval-wise testing for functional data.
\newblock \emph{Journal of Nonparametric Statistics}, 29\penalty0 (2):\penalty0
  407--424, 2017.

\bibitem[{R Development Core Team}(2022)]{rdevelopmentcoreteam2016}
{R Development Core Team}.
\newblock \emph{R: {{A Language}} and {{Environment}} for {{Statistical
  Computing}}}.
\newblock {R Foundation for Statistical Computing}, Vienna, Austria, 2022.

\bibitem[Ramsay and Silverman(2006)]{ramsay2005}
James~O. Ramsay and B.~W. Silverman.
\newblock \emph{Functional {{Data Analysis}}}.
\newblock Springer, New York, 2006.

\bibitem[Ramsay and Silverman(2002)]{ramsay2002}
James~O Ramsay and Bernard~W Silverman.
\newblock \emph{Applied {{Functional Data Analysis}}: {{Methods}} and {{Case
  Studies}}}.
\newblock Springer, New York, 2002.

\bibitem[Rue et~al.(2009)Rue, Martino, and Chopin]{rue2009}
H{\aa}vard Rue, Sara Martino, and Nicolas Chopin.
\newblock Approximate {Bayesian} inference for latent {Gaussian} models by
  using integrated nested {Laplace} approximations.
\newblock \emph{Journal of the Royal Statistical Society: Series B (Statistical
  Methodology)}, 71\penalty0 (2):\penalty0 319--392, 2009.

\bibitem[Rue et~al.(2017)Rue, Riebler, S{\o}rbye, Illian, Simpson, and
  Lindgren]{rue2017}
H{\aa}vard Rue, Andrea Riebler, Sigrunn~H S{\o}rbye, Janine~B Illian, Daniel~P
  Simpson, and Finn~K Lindgren.
\newblock Bayesian computing with inla: A review.
\newblock \emph{Annual Review of Statistics and Its Application}, 4:\penalty0
  395--421, 2017.

\bibitem[Schneider(2014)]{schneider2014}
Rolf Schneider.
\newblock \emph{Convex Bodies: The Brunn--Minkowski Theory}.
\newblock Cambridge University Press, Cambridge, 2014.

\bibitem[Schneider and Weil(2008)]{schneider2008}
Rolf Schneider and Wolfgang Weil.
\newblock \emph{Stochastic and Integral Geometry}.
\newblock Springer, New York, 2008.

\bibitem[Sepp{\"a} et~al.(2019)Sepp{\"a}, Rue, Hakulinen, L{\"a}{\"a}r{\"a},
  Sillanp{\"a}{\"a}, and Pitk{\"a}niemi]{seppa2019}
Karri Sepp{\"a}, H{\aa}vard Rue, Timo Hakulinen, Esa L{\"a}{\"a}r{\"a}, Mikko~J
  Sillanp{\"a}{\"a}, and Janne Pitk{\"a}niemi.
\newblock Estimating multilevel regional variation in excess mortality of
  cancer patients using integrated nested {L}aplace approximation.
\newblock \emph{Statistics in Medicine}, 38\penalty0 (5):\penalty0 778--791,
  2019.

\bibitem[Simpson et~al.(2016)Simpson, Illian, Lindgren, S{\o}rbye, and
  Rue]{simpson2016}
Daniel Simpson, Janine~Baerbel Illian, Finn Lindgren, Sigrunn~H S{\o}rbye, and
  Havard Rue.
\newblock Going off grid: Computationally efficient inference for
  log-{G}aussian {C}ox processes.
\newblock \emph{Biometrika}, 103\penalty0 (1):\penalty0 49--70, 2016.

\bibitem[Sokolova et~al.(2006)Sokolova, Japkowicz, and
  Szpakowicz]{sokolova2006}
Marina Sokolova, Nathalie Japkowicz, and Stan Szpakowicz.
\newblock Beyond accuracy, {F}-score and {ROC}: A family of discriminant
  measures for performance evaluation.
\newblock In \emph{Australasian Joint Conference on Artificial Intelligence},
  pages 1015--1021. Springer, 2006.

\bibitem[Spiegelhalter et~al.(2002)Spiegelhalter, Best, Carlin, and Van
  Der~Linde]{spiegelhalter2002}
David~J Spiegelhalter, Nicola~G Best, Bradley~P Carlin, and Angelika Van
  Der~Linde.
\newblock Bayesian measures of model complexity and fit.
\newblock \emph{Journal of the Royal Statistical Society: Series B (Statistical
  Methodology)}, 64\penalty0 (4):\penalty0 583--639, 2002.

\bibitem[Spiegelhalter et~al.(2014)Spiegelhalter, Best, Carlin, and Van~der
  Linde]{spiegelhalter2014}
David~J Spiegelhalter, Nicola~G Best, Bradley~P Carlin, and Angelika Van~der
  Linde.
\newblock The deviance information criterion: 12 years on.
\newblock \emph{Journal of the Royal Statistical Society: Series B (Statistical
  Methodology)}, 76\penalty0 (3):\penalty0 485--493, 2014.

\bibitem[Tao(2011)]{tao2011}
Terence Tao.
\newblock \emph{An Introduction to Measure Theory}.
\newblock American Mathematical Society, Providence, RI, 2011.

\bibitem[Waldmann(2014)]{waldmann2014}
Stefan Waldmann.
\newblock \emph{Topology: An Introduction}.
\newblock Springer, New York, 2014.

\bibitem[Wang et~al.(2018)Wang, Yue, and Faraway]{wang2018}
Xiaofeng Wang, Yuryan Yue, and Julian~J Faraway.
\newblock \emph{Bayesian regression modeling with INLA}.
\newblock Chapman and Hall/CRC, Boca Raton, FL, 2018.

\bibitem[Wu et~al.(2019)Wu, Zhang, and Du]{wu2019}
Wei-Li Wu, Zhao Zhang, and Ding-Zhu Du.
\newblock Set function optimization.
\newblock \emph{Journal of the Operations Research Society of China},
  7\penalty0 (2):\penalty0 183--193, 2019.

\bibitem[Xu et~al.(2020)Xu, Wang, Bian, Huang, Burch, Andrade, Zhang, and
  Guan]{xu2020}
Ganggang Xu, Ming Wang, Jiangze Bian, Hui Huang, Timothy~R. Burch, Sandro~C.
  Andrade, Jingfei Zhang, and Yongtao Guan.
\newblock Semi-parametric learning of structured temporal point processes.
\newblock \emph{Journal of Machine Learning Research}, 21\penalty0
  (192):\penalty0 1--39, 2020.

\bibitem[Young and Smith(2005)]{young2005}
G.~A Young and Richard~L Smith.
\newblock \emph{Essentials of {{Statistical Inference}}}.
\newblock {Cambridge University Press}, Cambridge, UK, 2005.

\end{thebibliography}

\end{document}